\documentclass[useAMS,usenatbib]{mn2e}
\usepackage{natbib}
\usepackage{amsmath,amsfonts}
\usepackage{epsfig}
\usepackage{lscape}
\def\reff@jnl#1{{\rm#1\/}}

\def\aj{\reff@jnl{AJ}}                  % Astronomical Journal
\def\araa{\reff@jnl{ARA\&A}}            % Annual Review of Astron and Astrophys
\def\apj{\reff@jnl{ApJ}}                        % Astrophysical Journal
\def\apjl{\reff@jnl{ApJ}}               % Astrophysical Journal, Letters
\def\apjs{\reff@jnl{ApJS}}              % Astrophysical Journal, Supplement
\def\ao{\reff@jnl{Appl.Optics}}         % Applied Optics
\def\apss{\reff@jnl{Ap\&SS}}            % Astrophysics and Space Science
\def\aap{\reff@jnl{A\&A}}               % Astronomy and Astrophysics
\def\aapr{\reff@jnl{A\&A~Rev.}}         % Astronomy and Astrophysics Reviews
\def\aaps{\reff@jnl{A\&AS}}             % Astronomy and Astrophysics, Supplement
\def\azh{\reff@jnl{AZh}}                        % Astronomicheskii Zhurnal
\def\baas{\reff@jnl{BAAS}}              % Bulletin of the AAS
\def\jrasc{\reff@jnl{JRASC}}            % Journal of the RAS of Canada
\def\memras{\reff@jnl{MmRAS}}           % Memoirs of the RAS
\def\mnras{\reff@jnl{MNRAS}}            % Monthly Notices of the RAS
\def\pra{\reff@jnl{Phys. Rev. A}}         % Physical Review A: General Physics
\def\prb{\reff@jnl{Phys. Rev. B}}         % Physical Review B: Solid State
\def\prc{\reff@jnl{Phys. Rev. C}}         % Physical Review C
\def\prd{\reff@jnl{Phys. Rev. D}}         % Physical Review D
\def\prl{\reff@jnl{Phys. Rev. Lett}}      % Physical Review Letter
\def\pasp{\reff@jnl{PASP}}              % Publications of the ASP
\def\pasj{\reff@jnl{PASJ}}              % Publications of the ASJ
\def\qjras{\reff@jnl{QJRAS}}            % Quarterly Journal of the RAS
\def\skytel{\reff@jnl{S\&T}}            % Sky and Telescope
\def\solphys{\reff@jnl{Solar~Phys.}}    % Solar Physics
\def\sovast{\reff@jnl{Soviet~Ast.}}     % Soviet Astronomy
\def\ssr{\reff@jnl{Space~Sci.Rev.}}     % Space Science Reviews
\def\zap{\reff@jnl{ZAp}}                        % Zeitschrift fuer Astrophysik
\def\nat{\reff@jnl{Nature}}             % Nature 

\def\p#1by#2{{\partial{#1} \over \partial{#2}}}
\def\pp#1by#2#3{{\partial^2{#1} \over \partial{#2}\partial{#3}}}
\def\d#1by#2{{{\rm d}{#1} \over {\rm d}{#2}}}
\def\dd#1by#2#3{{{\rm d}^2{#1} \over {\rm d}{#2}{\rm d}{#3}}}

\title[AMI limits on 15\,GHz excess emission in northern {\sc Hii} regions]
{AMI limits on 15\,GHz excess emission in northern {\sc Hii} regions}

\label{firstpage}
\author[Scaife et~al.]
{AMI CONSORTIUM: 
 Scaife A. M. M.$\thanks{E-mail: 
as595@mrao.cam.ac.uk}$, 
 Hurley-Walker N.,
 Davies, M. L., 
\newauthor
 Duffett-Smith, P. J.,
 Feroz, F., 
 Grainge K. J. B., 
 Green D. A.,
 Hobson M. P.,
\newauthor
 Kaneko, T.,
 Lasenby, A. N.,
 Pooley G. G.,
 Saunders R. D. E.,
 Scott P. F., 
\newauthor
 Titterington D. J.,
 Waldram E. M.,
 Zwart J.\\
 \vspace{0.03in}\\
Astrophysics Group, Cavendish Laboratory, 19 J. J. Thomson Avenue,
Cambridge CB3 0HE, UK.\\
}

\date{Accepted ---; received ---; in original form \today}
\pagerange{\pageref{firstpage}--\pageref{lastpage}}
\pubyear{2005}

\begin{document}
%\label{firstpage}
\maketitle

\begin{abstract}
We present observations between 14.2 and 17.9\,GHz of sixteen Galactic
{\sc Hii} regions made with the Arcminute Microkelvin Imager (AMI). In
conjunction with data from the literature at lower radio frequencies
we investigate the possibility of a spinning dust component in the
spectra of these objects. We conclude that there is no significant evidence for
spinning dust towards these sources and measure an average spectral
index of $\alpha = 0.15\pm0.07$ (where $S\propto\nu^{-\alpha}$) between 1.4 and 17.9\,GHz for the sample.

\end{abstract}

\begin{keywords}
ISM:{\sc Hii} regions -- radio continuum:ISM -- ISM:clouds -- radiation mechanisms:thermal
\end{keywords}

\section{Introduction}

Recent observations of Galactic targets (Finkbeiner et~al. 2002; 2004; Casassus
et~al. 2004;2006; Watson et~al. 2005; Scaife et~al. 2006; Dickinson et~al. 2007)
have provided some evidence for the anomalous microwave
emission commonly ascribed to spinning dust (Drain \& Lazarian
1998a,b). This emission was first seen as a large scale phenomenon in
CMB observations and represented a problem as it emits in the
frequency range ~ 10--60\,GHz (Kogut et~al. 1996; Leitch et~al. 1997; de
Oliviera-Costa et~al. 2002; 2004) close to the minimum of the combined
synchrotron, Bremsstrahlung and thermal dust emissions at
70\,GHz. Indeed the models of Draine \& Lazarian predict a spectrum
for spinning dust which is strongly peaked between 20 and
40\,GHz. Arising as a consequence of rapidly rotating small dust
grains the emission has been suggested to occur in a number of
distinct astronomical objects and to be highly correlated with thermal
dust emission; this has been supported by some of the pointed observations
referred to earlier and by recent evidence of diffuse emission at
medium to high galactic latitudes from correlations made with WMAP
data (Davies et~al. 2006).

Previous observations of {\sc Hii} regions in the microwave region of
the spectrum have shown evidence both for (Watson et~al. 2005;
Dickinson et~al. 2007) and against (Dickinson et~al. 2006; Scaife
et~al. 2007) the presence of a spinning dust component in the emission
of these objects. Since the behaviour of {\sc Hii} regions at radio
frequencies is relatively well understood they provide an excellent
testing ground for examining this phenomenon. At frequencies below
$\sim$100\,GHz {\sc Hii} regions are expected to be dominated by
free-free emission, or thermal Bremsstrahlung. This mechanism
progresses from the optically thick regime to the optically thin at
around 1\,GHz, and possesses a characteristically shallow
spectrum ($\alpha=0.1$; where $S\propto\nu^{-\alpha}$) at
frequencies above this turn over. 

Here we present observations of sixteen Galactic {\sc Hii} regions
selected from the VLA survey of optically visible Galactic {\sc Hii}
regions (Fich 1993). Using spectral data from the AMI in conjunction
with measurements from the literature we model the spectrum of
free-free radiation and compare it with our own measurements
in order to place limits on possible excess emission at 15\,GHz, which
may arise from spinning dust. We use the correlated FIR (100, 60, 25
and 12\,$\mu$m)
emission to further constrain this excess. 

\section{The Telescope}
The Arcminute Microkelvin Imager (AMI) is located at the Mullard Radio
Astronomy Observatory, Lord's Bridge, Cambridge, UK. Its Small Array is composed
of ten 3.7\,m diameter equatorially mounted dishes with a baseline range of
$\sim$5 to 20\,m. The telescope observes in the band 12--18\,GHz with cryostatically
cooled NRAO indium-phosphide front-end amplifiers. The system temperature
is typically about 25\,K. The astronomical signal is mixed with a
24\,GHz oscillating signal to produce an IF signal of
6--12\,GHz. The correlator is an analogue Fourier transform spectrometer
with 16 correlations formed for each baseline at path delays spaced
by 25\,mm. Both in phase and out of phase correlations are performed. From these, eight
channels of 750\,MHz bandwidth are synthesised. The AMI Small Array is sensitive to
angular scales of $\sim 1'$ to $\sim 15'$ and has a primary beam FWHM
of $\approx 20'$ at 15\,GHz. The lowest two channels are generally unused due
to a low response in this frequency range,
and strong interference from European geostationary satellites.

\section{Observations}

Observations of sixteen {\sc Hii} regions, see Table~\ref{tab:srclist}, were made with the AMI Small
Array during the period May--June 2007. These targets were selected
from the VLA survey of optically visible Galactic {\sc Hii} regions
(Fich 1993) on the basis of flux density, angular diameter and
declination. The sample, see Table~\ref{tab:srclist}, was divided on
the basis of flux at 4.89\,GHz. The positions and effective thermal
noise associated with each AMI 
observation are shown in Table~\ref{tab:srclist}. 
Observations
were typically 8 hours long and used interleaved calibration on bright
point sources at hourly intervals for phase calibration. The sensitivity of AMI is $\approx$30\,mJy\,s$^{-1/2}$
giving $\approx$0.2\,mJy after 8 hours observation. Given
that the objects observed here are bright, eight hours is more than
sufficient to obtain accurate fluxes; however the
long observations are important to produce good maps, the structure
of which is dependent on the {\it uv}-coverage. 
\begin{table}
\centering
\caption{AMI {\sc Hii} region sample. Names of the form S$nnn$ are taken from the
  Sharpless (1959) catalogue of {\sc Hii} regions; and names of the
  form BFS$nn$ are taken from the catalogue of Blitz, Fich \& Stark
  (1982). Radio positions are 
  adapted from Fich (1993). Column 4 contains the effective thermal noise
  for the AMI observation of each object.\vspace{0.2cm}\label{tab:srclist}} 
\begin{tabular}{lccc}
\hline\hline
Name & $\alpha^{1}$ & $\delta$ & $\sigma_{\rm{th}}$ \\
&(J2000)&(J2000)&(mJy)\\
\hline
\hline
\multicolumn{4}{l}{$S_{4.89}$ $\ge$ 1\,Jy:}\\
\hline
S100&	20 01 44.7&	+33 31 14 & 7.760  \\	
S152&	22 58 40.8&	+58 47 02 & 1.099 \\	
\hline
\multicolumn{4}{l}{$S_{4.89}$ $\ge$ 0.5\,Jy:}\\
\hline
S127&	21 28 38.0&	+54 35 12 & 0.710 \\	
S138&	22 32 45.2&	+58 28 21 & 0.629 \\	
S149&	22 56 17.4&	+58 31 18 & 0.593 \\	
S211&	04 36 57.0&	+50 52 36 & 0.646 \\	
S288&	07 08 37.0&	-04 18 48 & 3.703 \\	
\hline
\multicolumn{4}{l}{$S_{4.89}$ $\ge$ 0.1\,Jy:}\\
\hline
S121&	21 05 15.8&	+49 40 06 & 0.650 \\	
S167&	23 35 30.9&	+64 52 28 & 0.330 \\	
S175&	00 27 17.0&	+64 42 23 & 0.319 \\	
S186&	01 08 50.5&	+63 07 34 & 1.033 \\	
S256&	06 12 36.0&	+17 56 54 & 1.288 \\	
S259&	06 11 25.8&	+17 26 25 & 0.537 \\	 
S271&	06 14 59.4&	+12 20 16 & 0.652 \\	
BFS10&	21 56 30.4&	+58 01 43 & 0.501 \\	 
BFS46&	05 40 52.9&	+35 42 17 & 0.727 \\	

\hline
\end{tabular}
\begin{minipage}{7cm}
{\small\vspace{0.1cm} $^{1}$ positions adapted from Fich (1993)} 
\end{minipage}
\end{table}

\begin{table}
\centering
\caption{\label{tab:Fluxes-of-3C286}Assumed fluxes of 3C286 and 3C48 over the AMI
bandwidth.}
\begin{tabular}{|c|c|c|c|}
\hline 
Channel&
$\bar{\nu}$&
$S^{\rm{3C286}}$/Jy&
$S^{\rm{3C48}}$/Jy\tabularnewline
\hline
\hline 
1&
12.788&
3.909&
1.941\tabularnewline
2&
13.512&
3.755&
1.832\tabularnewline
3&
14.235&
3.614&
1.734\tabularnewline
4&
14.958&
3.485&
1.646\tabularnewline
5&
15.682&
3.366&
1.566\tabularnewline
6&
16.405&
3.257&
1.494\tabularnewline
7&
17.128&
3.155&
1.428\tabularnewline
8&
17.852&
3.061&
1.367\tabularnewline
\hline
\end{tabular}
\end{table}

\begin{table*}
\centering
\caption{Phase calibrators used during
{\sc Hii} region observations. 8\,GHz flux densities are from JVAS; 15\,GHz
flux densities are from the VLA Calibration Source List$^{1}$ and from the
AMI {\sc Hii} region sample described in this work. If a source is
known to variable it is indicated.\label{tab:Phase-calibrators-used}}
\begin{tabular}{|l|c|c|c|c|c|c|c|}
\hline 
Source&
$\alpha$(J2000)&
$\delta$(J2000)&
$S_{8}$(Jy)&$S_{15}^{\rm{VLA}}$(Jy)&$S_{15}^{\rm{AMI}}$(Jy)&variable?&
{\sc Hii} Regions calibrated\tabularnewline
\hline
\hline 
J0100+681&
01 00 51.7&
+68 08 21&
0.856&
0.6&
0.610&
no&
S167, S175\tabularnewline
J0102+584&
01 02 45.8 &
+58 24 11&
1.399&
2.3&
3.070&
yes&
S186\tabularnewline
J0359+509&
03 59 29.7&
+50 57 50.2&
2.404&
10.5&
9.280&
yes&
S211\tabularnewline
J0539+1433&
05 39 42.4&
+14 33 46&
1.093&
0.5&
0.660&
yes&
S256, S259, S271\tabularnewline
J0555+398&
05 55 30.8&
+39 48 49&
6.984&
2.8&
3.560&
yes&
BFS46\tabularnewline
J0739+0137&
07 39 18.0&
+01 37 05&
1.746&
2.1&
2.430&
yes&
S288\tabularnewline
J2025+3343&
20 25 10.8&
+33 43 00&
2.592&
2.5&
2.410&
yes&
S100\tabularnewline
J2038+513&
20 38 37.0&
+51 19 13&
4.205&
3.3&
3.140&
yes&
S121\tabularnewline
J2125+643&
21 25 27.4&
+64 23 39&
1.127&
0.6&
0.570&
no&
BFS10\tabularnewline
J2201+508&
22 01 43.5&
+50 48 56&
0.815&
0.6&
0.300&
yes&
S127\tabularnewline
J2322+509&
23 22 26.0&
+50 57 52&
1.656&
1.6&
0.940&
yes&
S152, S138, S149\tabularnewline
\hline
\end{tabular}
\begin{minipage}{16cm}
{\small $^{1}$Version (08/2002): {\tt http://www.nrao.edu/$\sim$gtaylor/calib.html}}
\end{minipage}
\end{table*}

\section{Calibration and Data reduction}

Data reduction was performed using the local software tool {\sc reduce},
developed from the VSA data-reduction software of the same name. This
applies appropriate path compensator and path delay corrections, flags
interference, shadowing and hardware errors, applies phase and flux
calibrations and Fourier transforms the data into {\it uv} {\sc fits} 
format suitable for imaging in {\sc aips}.

Flux calibration was performed using short observations of 3C48 and
3C286 near the beginning and end of each run; assumed flux densities
for these sources in the AMI channels as taken from Baars
et~al. (1977) are shown in 
Table.~\ref{tab:Fluxes-of-3C286}. In addition to the flux calibration
each observation was interleaved with a secondary phase calibrator. 

Secondary calibrators are selected from the Jodrell Bank
VLA Survey (hereinafter JVAS; Patnaik et~al. 1992; Browne et~al. 1998; Wilkinson
et~al. 1998) on the basis of their declination and flux. A list
of these calibrators is given in Table \ref{tab:Phase-calibrators-used}.
Over one hour the phase is generally stable to $5^{\circ}$ for Channels 4--7, and
$10^{\circ}$ for Channels 3 and 8. 

A weather correction is also made using data from the `rain gauge'.
This measures the sky temperature in correlator units which are calibrated
by measuring the rain gauge response on a cool, dry, clear day. From
cross-calibration of our primary calibrators we
find that the weather correction and primary calibration give fluxes
correct within 5 per cent.

Low declination observations can be highly contaminated by geostationary
satellites. Much of this emission is in the low frequency channels,
which are usually discarded. However narrow-band emission at 15\,GHz
can seriously contaminate astronomical data. As the satellites are
fixed against the moving astronomical sky, a high-pass filter applied
to the phase centre before fringe rotation can remove the bulk of
the contamination, although this results in the loss of all low-{\it v}
visibilities in the {\it uv}-plane. 

\section{Imaging and Spectra}

Reduced data were imaged using the {\sc aips} data package. Maps were made
from both the combined channel set, shown in this paper, and from individual
channels. The broad spectral coverage of the AMI allows a
representation of the spectrum between 14 and 18\,GHz to be
made. Errors on the AMI data points were 
calculated using a 5 per cent error on the flux and the thermal noise
of each individual observation as calculated from the data. This error
of 5 per cent on the flux of each source is a conservative error on
the day-to-day calibration of the telescope which has been found to be
significantly better than 5 per cent. It also includes a contribution
for the Gaussian fitting of the sources although in most cases this
fitting was found to be robust to changes in fitting area, a test
which usually reveals cases where the source is poorly fitted. The overall
error was calculated as $\sigma =
\sqrt{(\sigma_{\rm{th}}^2+(0.05S_{\rm{i}})^2)}$, where
$\sigma_{\rm{th}}$ is the thermal noise calculated outside the primary
beam for that observation and $S_{\rm{i}}$ is the integrated flux density of the source. For those
observations heavily contaminated by satellite interference a more
conservative 10 per cent error is placed on the flux density
calibration. The central frequency of channels 3--8 is 15.8\,GHz and
fluxes from the combined channels can be found in
Table~\ref{tab:results}.

In the Northern sky a number of
Galactic surveys exist. At 1.4\,GHz,
measurements from the Effelsberg telescope (Reich et~al. 1997) and
from the NVSS (Condon et~al. 1998) can be compared, with due caution employed
considering the relative angular resolution and inherent flux losses
from the VLA compared with the Effelsberg 100\,m dish. In addition to
these the Canadian Galactic Plane Survey (hereinafter CGPS) 
provides pseudo total power observations of the Galactic plane at 1.4\,GHz
by combining single dish and interferometer data to achieve a resolution of 1\,arcmin. At 2.7\,GHz the
Effelsberg Galactic plane survey (F\"{u}rst et~al. 1990a;b)is available,
and the low
latitude Galactic survey from the Parkes observatory (Day
et~al. 1972) also covers a small number of our low declination sources. At higher frequencies the GB6 survey
(Gregory et~al. 1996) at
4.85\,GHz and the Galactic plane survey of Langston et~al (2000) at
8.35 and 14.35\,GHz can be used, although the detection limit of
0.9\,Jy at 8.35\,GHz 
and 2.5\,Jy at 14.35\,GHz restricts these measurements to
only the brightest sources.

In addition to these surveys a number of pointed {\sc Hii} samples
have been made including those of Gregory and Taylor (1986) at 5\,GHz,
Kallas \& Reich (1980); Kazes,
Walmsley \& Churchwell (1977) again with the Effelsberg 100\,m dish,
Felli (1978) at 6\,arcsecond resolution using the Westerbork telescope
and a number of small samples of {\sc Hii} regions were presented in a
series of papers (Israel, Habing \& de Jong 1973; Israel 1976a,b;
1977a,b,c).

\section{Results}
\label{sec:6}

\begin{figure}
\centerline{\includegraphics[height=7.cm,width=6.cm,angle=-90]{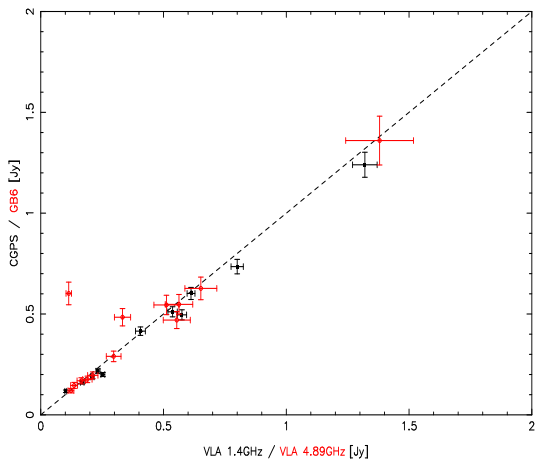}}
\caption{Correlation of VLA 1.4 and 4.89\,GHz flux densities with those from
  the Canadian Galactic Plane Survey at 1.4\,GHz and the GB6 catalogue at 4.85\,GHz. \label{fig:corr}}
\end{figure}

Gaussian fits were made to the sources found in the AMI images, allowing
for a background component. The integrated flux of each source was
then used in conjunction with fluxes from the literature to constrain
the radio spectrum of each object. Since the bandwidth of AMI
stretches from 14.2\,GHz to 17.9\,GHz it is possible to constrain
the microwave spectrum of each {\sc Hii} region independently of
other measurements. However, to investigate the spectral behaviour
of these objects through the radio and into the microwave regime we
also combine our own fluxes with those from other catalogues. We include
data from the NVSS catalogue at 1.4\,GHz and also from the VLA survey of optically
identified {\sc Hii} regions from which our sample is taken. The
second of these catalogues has measurements
at 4.89\,GHz, and occasionally 1.42\,GHz. Since no uncertainties are quoted for these flux
densities we adopt 
a conservative error of 10 per cent. These measurements have been shown to contain a
minimal amount of flux loss compared to single dish observations (Fich
1993),
i.e. they do not appear to resolve out flux on scales larger than those
measured;
we confirm this by comparing them to the GB6 survey which, although
not a total power measurement, contains information on significantly
larger scales than the 4.89\,GHz VLA data, which has an angular
resolution of 13\,arcsec compared with the 3.4\,arcmin resolution of
GB6. This comparison provides a robust assessment of the flux loss since the
angular scales measured by AMI lie between the two ranges. The result
of this comparison is shown in Figure~\ref{fig:corr} where a good
correlation can be seen within the errors in all but three
cases. These discrepant flux densities are S138, S121 and S256. In the
case of S121 and S256, where GB6 exhibits a significantly higher flux
density this effect can be attributed to the relative resolution of
the two instruments, with GB6 including adjacent point
sources which then contribute to the flux. In the case of S138, where
GB6 lists a flux density of 470$\pm$42\,mJy, the
cause of this discrepancy is less clear. However we note that the
4.85\,GHz radio catalogue of Becker, White and Edwards (1991) also made using the
Green Bank 91\,m dish records a flux density of 584$\pm$59\,mJy for
this source which is much
more consistent with that of the VLA at 554$\pm$55\,mJy. To assess
flux losses at 1.4\,GHz we compare NVSS data with a resolution of
45\,arcsec to data taken from the total power measurements of
the CGPS, which has a
resolution of 1\,arcmin. These data also show a tight correlation (see
Figure 1). The one source missing from this plot is S100, the high flux
density of which precludes it through necessities of scale, however
its flux density at 1.4\,GHz from the CGPS of 9.2\,Jy agrees well with
that from NVSS of $9.00\pm0.45$\,Jy.

These surveys provide data which we believe to be
reliable in relation to the fitting procedures used to determine flux
densities from the AMI data. We fit power
laws to the lower frequency radio data (NVSS 1.4 and VLA 4.89\,GHz), the
AMI data on its own and the combined data sets. The derived properties
of these spectra may be found in
Table~\ref{tab:results}, and are discussed in detail in Section 7. 

Since we are using interferometric data we can also fit spectra
directly to the visibility measurements in the Fourier plane, a method
which is independent of any subjective cleaning procedures. Where we
have observed isolated objects the
results of these fits agree closely 
with those found using the data in the map plane, and overall we find
there is a good agreement between the two methods. However the
complexity of modelling non-isolated extended sources in the {\it uv}
precludes us from performing all our measurements in this manner.

\subsection{Notes on Individual Sources}

In the following sections we will discuss the {\sc Hii} regions in
order of Right Ascension.\\

\begin{figure}
\centerline{\includegraphics[height=9.cm,width=9.cm,angle=0]{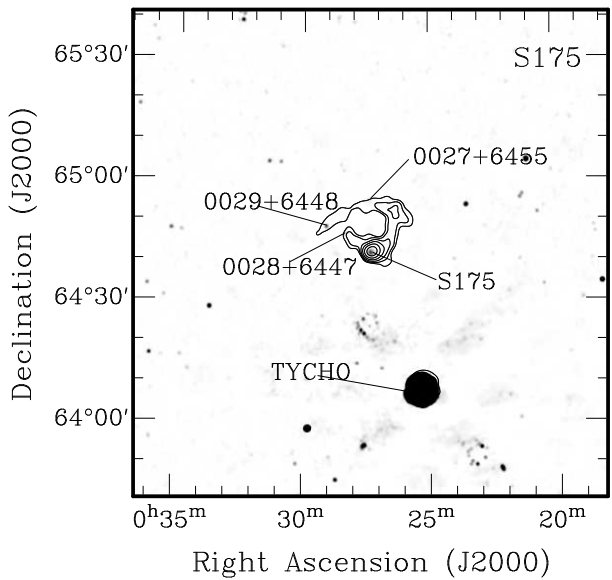}}
\centerline{\includegraphics[height=8.5cm,width=4.cm,angle=-90]{./S175_sp.ps}}
\caption{Above: Map of the S175 region. AMI 15.8\,GHz contours are overlaid
  on an NVSS 1.4\,GHz greyscale image. Contours increase in powers of
  2 from 4\,$\sigma$ (i.e. 4,8,16,32,64 etc). Below: Radio spectrum of S175. Data points are integrated flux
  densities taken from the literature, see
  Table~\ref{tab:radioflux}. A best fitting power law is shown as a dashed line. \label{fig:s175spec}}
\end{figure}

\begin{figure}
\centerline{\includegraphics[height=7.cm,width=7.cm,angle=0]{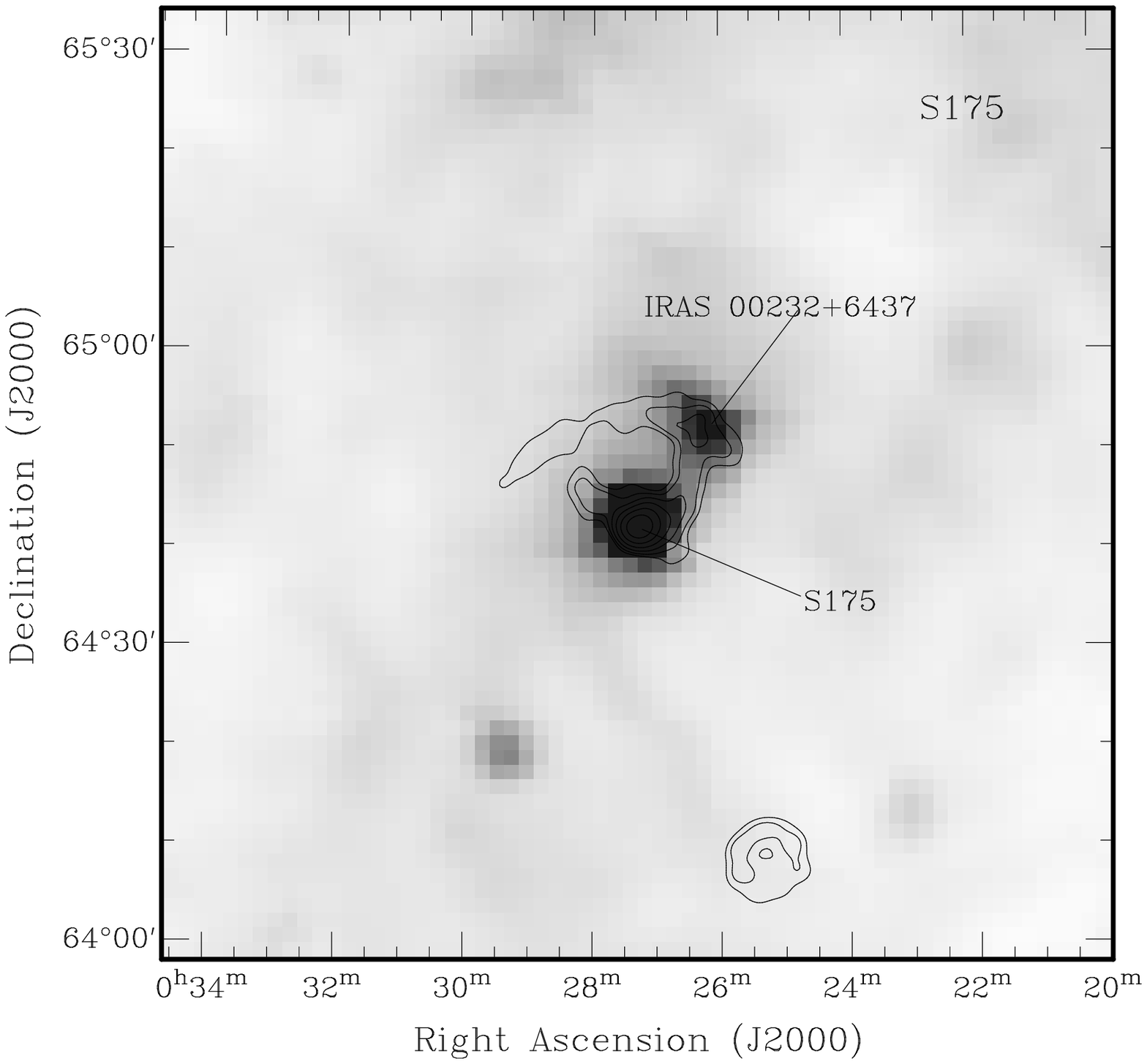}}
\caption{Map of the S\,175 region. AMI 15.8\,GHz contours are overlaid
  on an IRAS 100\,$\mu$m greyscale image. Contours are as in Figure~\ref{fig:s175spec}. \label{fig:s175iras}}
\end{figure}

\noindent
\textbf{S175} (Figures~\ref{fig:s175spec}
and~\ref{fig:s175iras}). The bipolar nebula S175 is seen at 15.8\,GHz to sit in a
ring of extended emission, the western side of which is visible in the
IRAS 100\,$\mu$m data, see Figure~\ref{fig:s175iras}, and the eastern side of which contains a number
of small radio sources visible at 1.4\,GHz in the NVSS survey. Also
present at radio wavelengths is the Tycho SNR (=G120.1+2.1), still visible at 15.8\,GHz
approximately one primary beam away from the pointing center. Fitting
a modified Planck spectrum of
the form $\nu^2B(\rm{T})$ (Lagache et~al. 2000) to the
IRAS 100/60\,$\mu$m flux densities we calculate a dust temperature of
$T_{\rm{d}} = 27.9$\,K towards S175. Using the optical recombination line measurements of Hunter
(1992) we determine the temperature of the electron gas using the
following equation (Haffner, Reynolds \& Tufte 1999):

\begin{equation}
\label{equ:nh}
\frac{I_{\rm{{\sc{NII}}}}}{I_{\rm{{\sc H\alpha}}}} = 1.63\times10^5\left(\frac{N}{{H}}\right)T_4^{0.426}\rm{e}^{-2.18/T_4},
\end{equation}

\noindent
where $T_4$ is the electron gas temperature in units of 10$^4$\,K. We use the solar ($N$/$H$) = $7.5\times10^{-5}$ from Meyer, Cardelli, \&
Sofia (1997) for the gas-phase abundance of N. This gives an electron
temperature of $T_{\rm{e}} = 7000\pm200$\,K.

The radio spectra of S175 is unusual with a gently climbing
spectral index of $\alpha = -0.14$ between 1.4 and 5\,GHz. It would be
convenient to be able to attribute this rising index to a combination of
resolution effects considering the much larger resolution of the GB6 and Effelsberg 100\,m dish, and flux
losses from the VLA at 1.4\,GHz if it were not for the continuation of
the index from the low resolution Effelsberg total power flux density at 2.7\,GHz to
the high resolution VLA interferometric flux density at 4.89\,GHz. The good agreement
of the GB6 data at 4.85\,GHz with the VLA at 4.89\,GHz also suggests
that there is no flux loss towards this object. A further interesting
feature of this spectrum is the apparent bend in the 15\,GHz data from
the AMI as the data flattens off towards higher frequencies. It is
most likely that the rising spectrum between 1.4 and 5\,GHz is due to
compact knots within the region which are optically thick for $\nu<5$\,GHz.\\

\begin{figure}
\centerline{\includegraphics[height=9.cm,width=9.cm,angle=0]{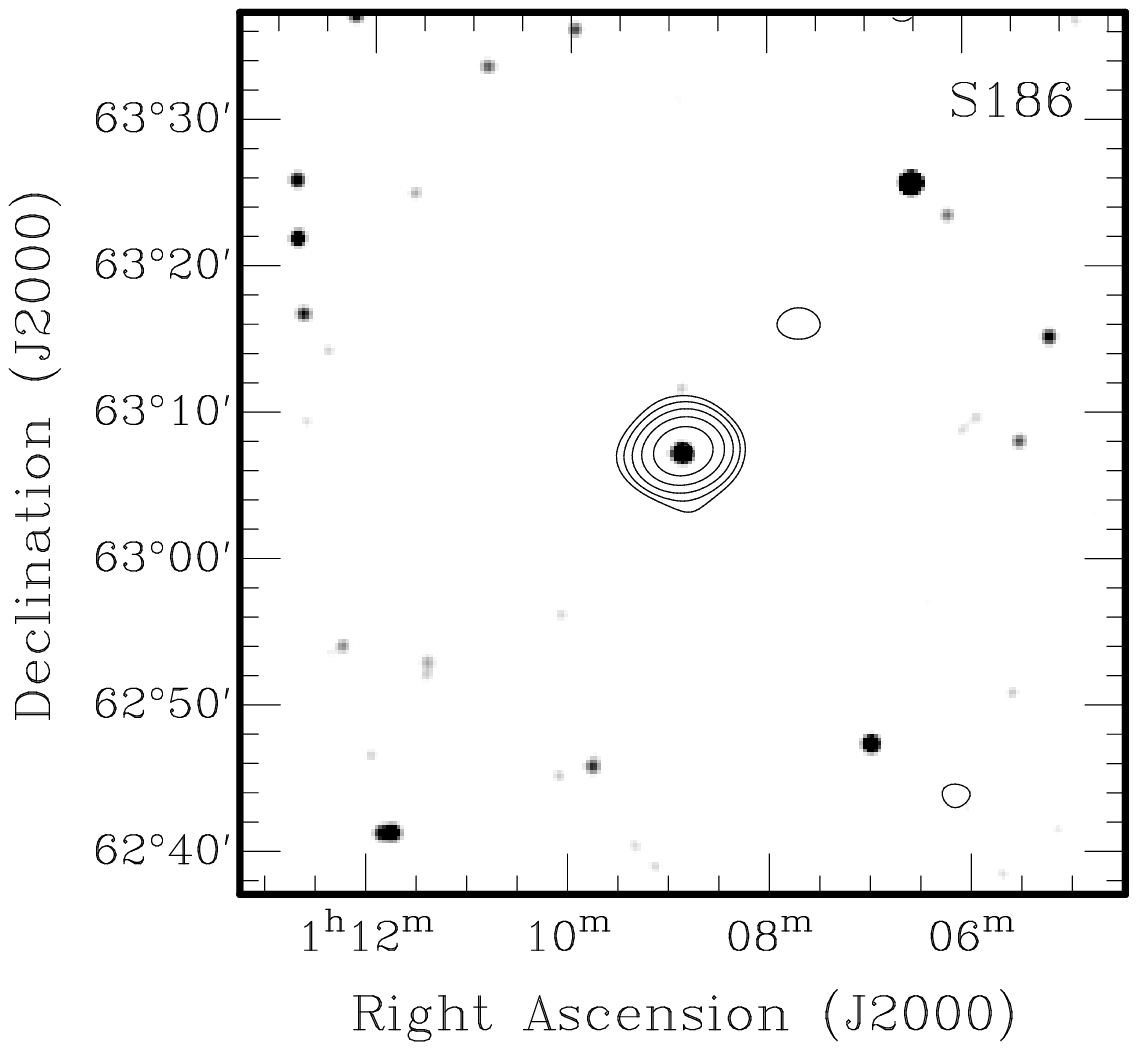}}
\centerline{\includegraphics[height=8.5cm,width=4.cm,angle=-90]{./S186_sp.ps}}
\caption{Above: Map of the S186 region. AMI 15.8\,GHz contours are overlaid
  on an NVSS 1.4\,GHz greyscale image. Contours are as in Figure~\ref{fig:s175spec}. Below: Radio spectrum of S186. Data points are integrated flux
  densities taken from the literature, see
  Table~\ref{tab:radioflux}. A best fitting power law is shown as a dashed line. \label{fig:s186spec}}
\end{figure}

\noindent
\textbf{S186} (Figure~\ref{fig:s186spec}) This object is a small nebulosity with a visible 100\,$\mu$m IRAS
association (IRAS 01056+6251). It is optically thin at low radio
frequencies with a flux density of 178\,mJy at 330\,MHz (Rengelink
et~al. 1997). From its IRAS 100/60\,$\mu$m flux densities we fit a
dust temperature of 33.2\,K and from optical recombination line data
(Hunter 1992) we calculate an electron gas temperature of
7300\,K (no uncertainty given). Although the spectral index across the AMI band is slightly
steep at $\alpha_{\rm{AMI}} = 0.23\pm1.34$ it is entirely consistent with
an overall index of $\alpha^{1.4}_{18} = 0.08\pm0.06$. \\

\begin{figure}
\centerline{\includegraphics[height=9.cm,width=10.cm,angle=0]{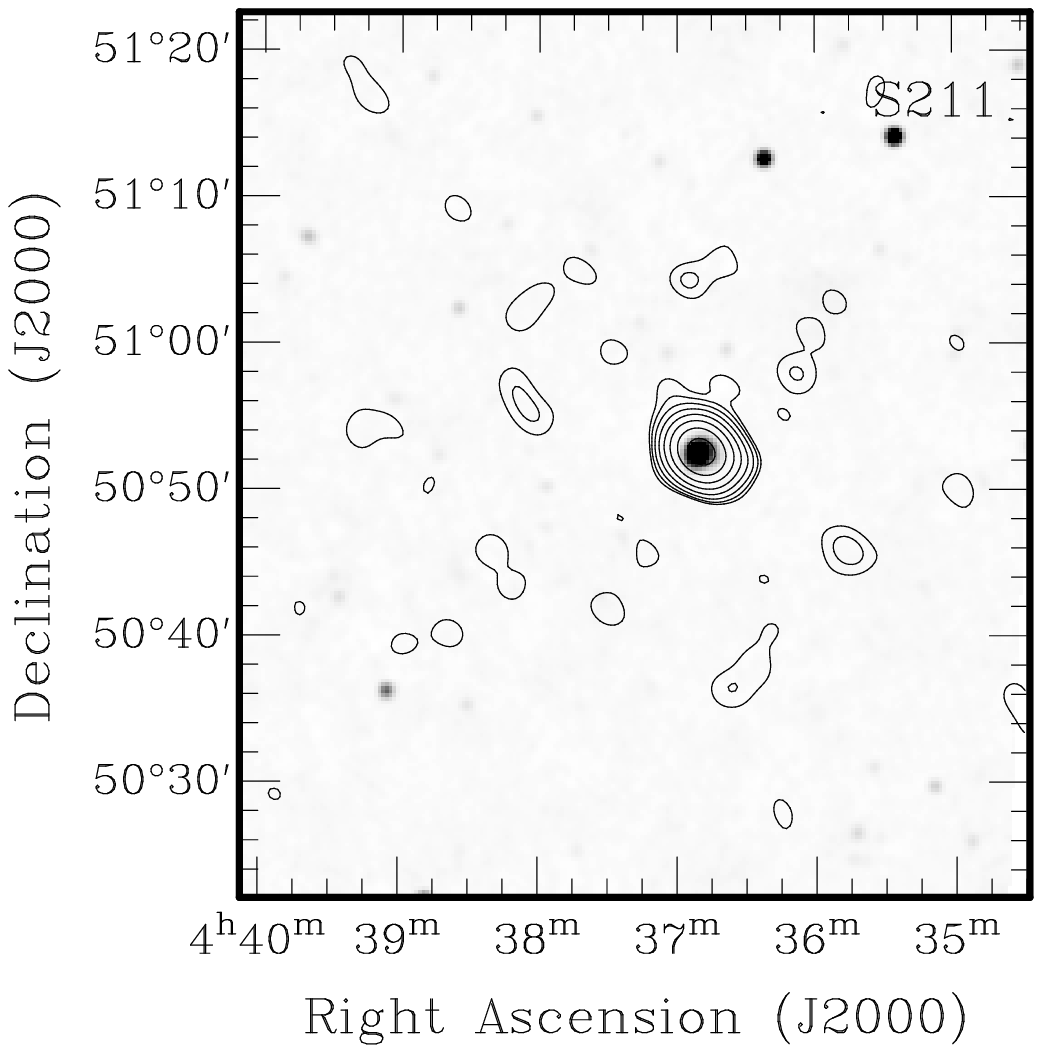}}
\centerline{\includegraphics[height=8.5cm,width=4.cm,angle=-90]{./S211_sp.ps}}
\caption{Above: Map of the S211 region. AMI 15.8\,GHz contours are overlaid
  on an NVSS 1.4\,GHz greyscale image. Contours are as in Figure~\ref{fig:s175spec}. Below: Radio spectrum of S211. Data points are integrated flux
  densities taken from the literature, see
  Table~\ref{tab:radioflux}. A best fitting power law is shown as a
  dashed line. \label{fig:s211spec}
}
\end{figure}

\noindent
\textbf{S211} (Figure~\ref{fig:s211spec}) This source, otherwise known as LBN\,717, is relatively isolated in the radio
although at high resolution (Fich 1993) shows a complex structure of
small knots embedded in a larger nebulous region. In the radio S211
has been observed at a number of frequencies. We include data at
1.4\,GHz from the NVSS 
and 4.89\,GHz from the VLA (Fich 1993); at 2.7\,GHz from the
Effelsberg 100\,m dish (F{\"u}rst et~al. 1999), which we believe to be useful here due
to the relatively isolated nature of this object; data at 4.85\,GHz
from the GB6 survey, the value of which agrees well with that of the
VLA at a different range of angular scales, illustrating the compact
nature of this source; and data at 3.2, 6.6 and 10.7\,GHz from the
Alonquin 46\,m dish (Andrew et~al. 1973).

The IRAS 100\,$\mu$m
data show a neighbouring region which we associate with IRAS
04330+5105, however at a distance of 18.5\,arcmin it is too far down the
primary beam to be detected by AMI. Fitting to the IRAS 100/60$\mu$m flux
densities we derive a dust temperature for this region of
31.4\,K and from the optical recombination line data of Hunter (1992)
we calculate an electron gas temperature of $T_{\rm{e}} = 7000\pm200$\,K. \\

\begin{figure}
\centerline{\includegraphics[height=9.cm,width=9.cm,angle=0]{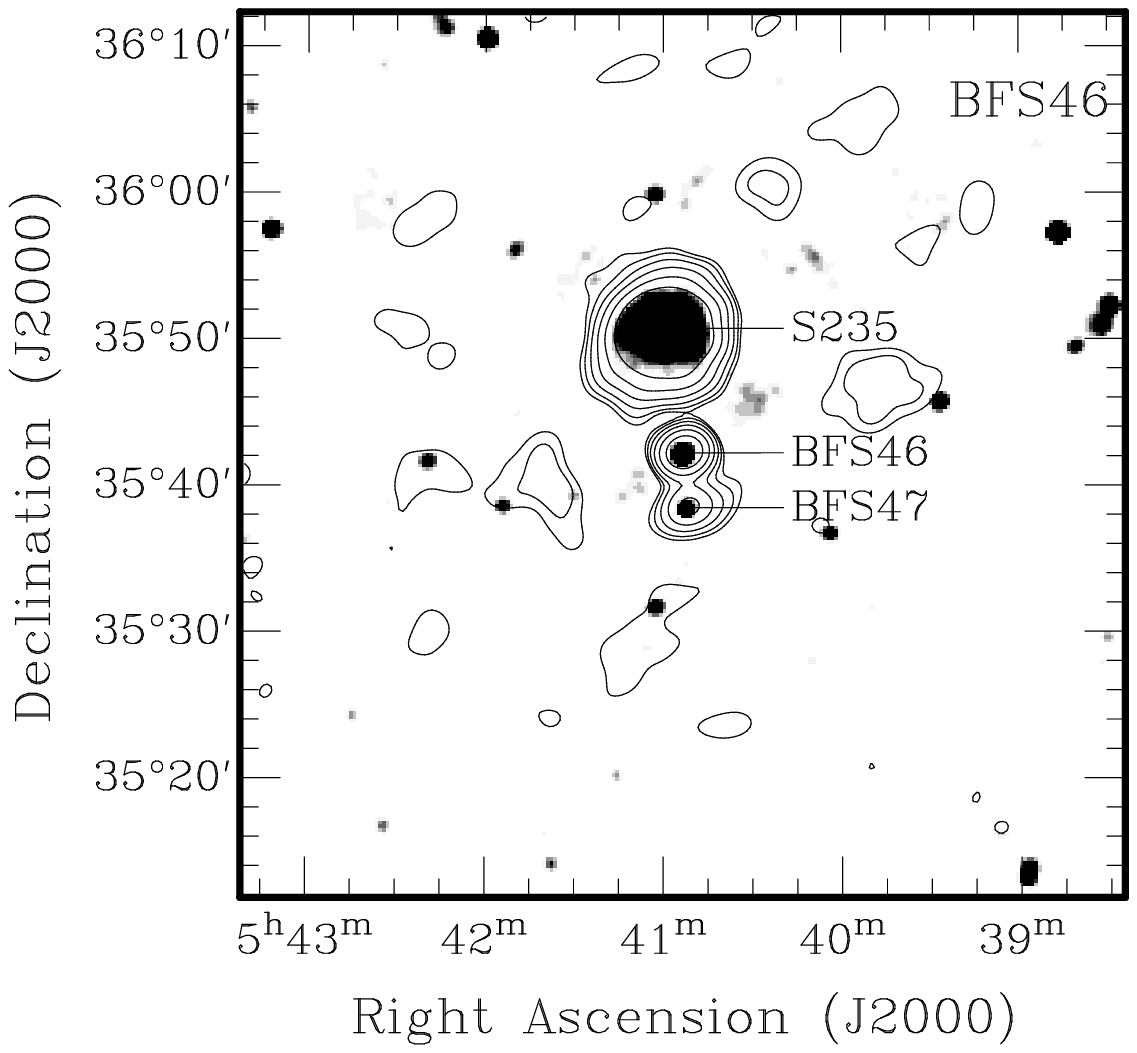}}
\centerline{\includegraphics[height=8.5cm,width=4.cm,angle=-90]{./BFS46_sp.ps}}
\caption{Above: Map of the BFS46 region. AMI 15.8\,GHz contours are overlaid
  on an NVSS 1.4\,GHz greyscale image. Contours are as in Figure~\ref{fig:s175spec}. Below: Radio spectrum of BFS46. Data points are integrated flux
  densities taken from the literature, see
  Table~\ref{tab:radioflux}. In addition data at 4.75, 8.45, 23 and
45\,GHz from the VLA are also shown. A best fitting power law is shown as a
  dashed line. \label{fig:bfs46spec}
}
\end{figure}

\noindent
\textbf{BFS46} (Figure~\ref{fig:bfs46spec}) Also called S235A (Felli et~al. 2004; 2006) owing to its
close proximity to the more extended S235 {\sc Hii} region (Sharpless
1959; Felli \& Churchwell 1972), BFS46 is a small region of
nebulosity. Its spectrum from 1.4 to 5\,GHz would suggest partially
thick emission due to its rising spectral index (Israel \& Felli 1978)
but at frequencies above 5\,GHz it appears optically thin. The GB6
flux density measurement at 4.85\,GHz is unusually low compared to
other values at similar frequencies and we suggest that this may be a
consequence of difficulty in fitting for a background component in such a
densely populated region. We also include data at 4.75, 8.45, 23 and
45\,GHz from the VLA (Felli et~al. 2006) although we note that the
fluxes at 23 and 45\,GHz are likely to be affected by flux
losses. The spectrum across the AMI band is consistent with optically
thin thermal emission, having a spectral index of $\alpha_{\rm{AMI}}
= 0.12\pm0.81$. Including VLA data at 1.4 and 4.89\,GHz gives an
overall spectral index of $\alpha = 0.09\pm0.04$.

Although we obtain a dust temperature of $T_{\rm{d}} = 38.5$\,K, 
data do not exist in the literature to calculate an electron gas
temperature for this object.\\  
\begin{figure}
\centerline{\includegraphics[height=8.cm,width=8.5cm,angle=0]{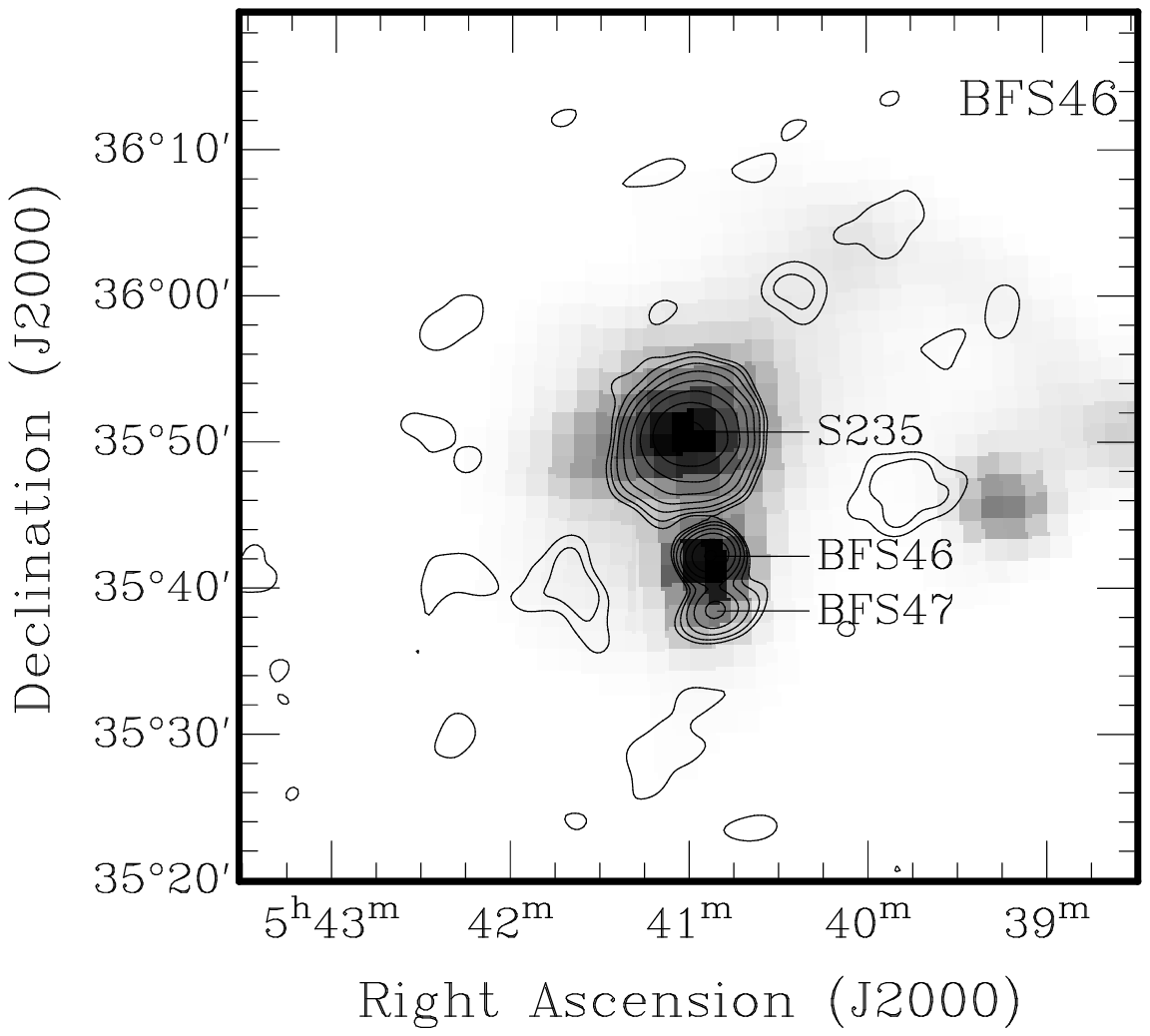}}
\caption{AMI 15.8\,GHz contours are overlaid
  on an IRAS 100\,$\mu$m greyscale image. Contours are as in Figure~\ref{fig:s175spec}. \label{fig:bfs46map}}
\end{figure}

\begin{figure}
\centerline{\includegraphics[height=9.cm,width=9.cm,angle=0]{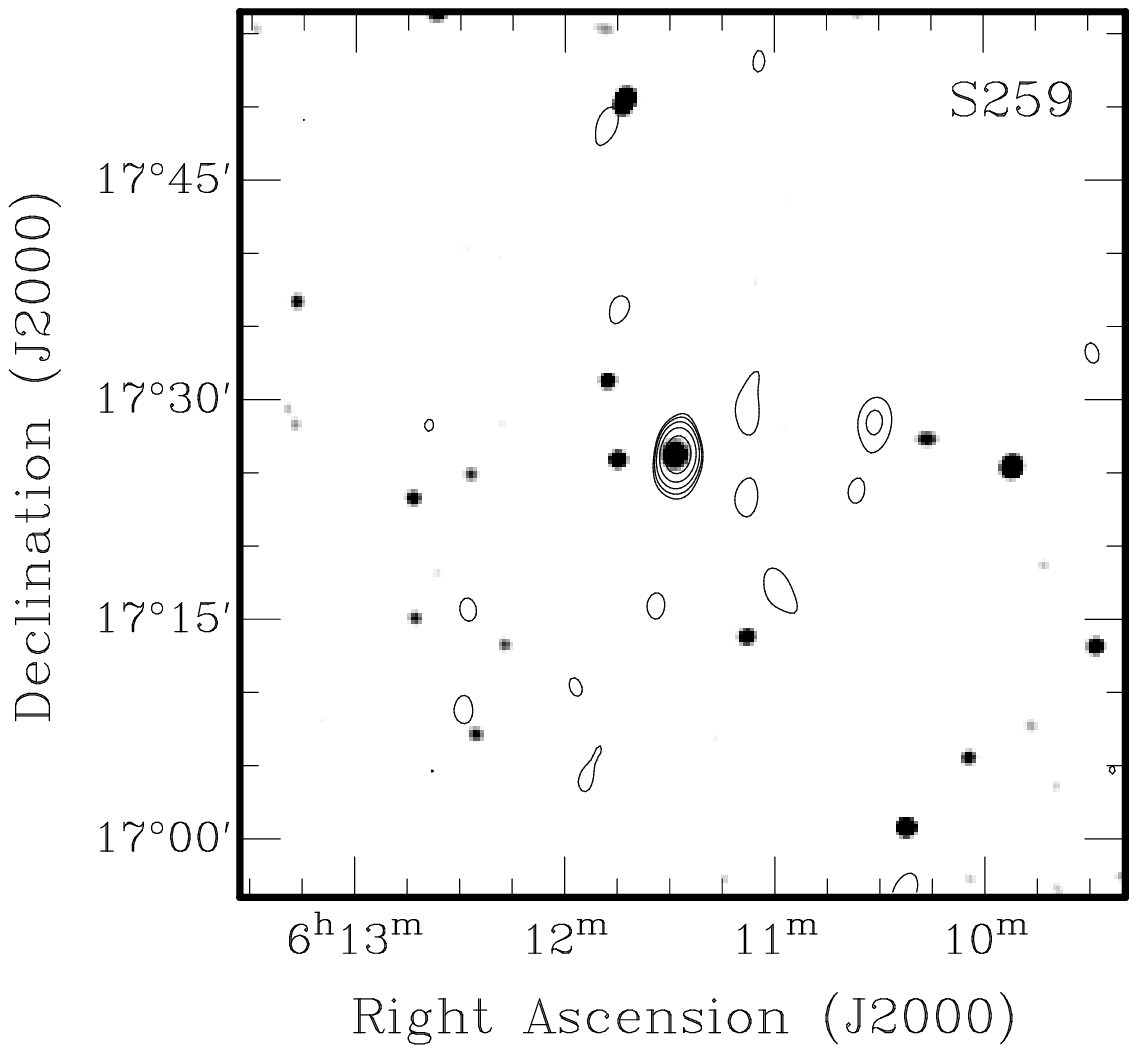}}
\centerline{\includegraphics[height=8.5cm,width=4.cm,angle=-90]{./S259_sp.ps}}
\caption{Above: Map of the S259 region. AMI 15.8\,GHz contours are overlaid
  on an NVSS 1.4\,GHz greyscale image. Contours are as in Figure~\ref{fig:s175spec}. Below: Radio spectrum of S259. Data points are integrated flux
  densities taken from the literature, see
  Table~\ref{tab:radioflux}. The best fitting power law is
  shown as a dashed line. \label{fig:s259spec}}
\end{figure}

\noindent
\textbf{S259} (Figure~\ref{fig:s259spec}) The isolated {\sc Hii} region S259 lies almost directly
south of the S254-257 complex. Although it is seen towards the Gemini
OB 1 molecular cloud complex it is presumed to be a background source
at a distance of 8.3\,kpc (Carpenter et~al. 1995). In the case of this object
the radio emission is far more compact than the IR with a partial ring
of IR emission seen to the west (Deharveng et~al. 2005) at shorter
wavelengths.

At a declination of $\sim17^{\circ}$ this observation is still heavily
affected by satellite interference in the AMI band and the spectral
index derived from these frequencies is not well constrained. 

Although data are not available in the literature to make an exact
calculation of the electron gas temperature of S259 we place an upper
limit on the temperature using the H$\alpha$ emission line measurements
of Fich, Treffers \& Dahl (1990). Assuming that the line width is due
largely to Doppler broadening we correct the line widths for the
filter response using the simple Gaussian approximation
$\Delta\nu_{\rm{D}} = \left[\Delta\nu_{\rm{L}}^2 -
  \Delta\nu_{\rm{f}}^2\right]^{1/2}$ (Reifenstein et~al. 1970). From
the corrected line widths we can combine
\begin{equation}
\Delta\nu = \frac{\nu_0\Delta V_{\rm FWHM}}{2 { c}\sqrt{2 {\rm ln}(2)}}
\end{equation}
and
\begin{equation}
\Delta\nu = \nu_0\sqrt{\frac{{ k}T_{\rm e}}{m{ c}^2}},
\end{equation}
to give
\begin{equation}
\label{equ:te}
T_{\rm e} = \frac{m}{ k} \left(\frac{\Delta V_{\rm FWHM}}{2\sqrt{2 {\rm ln}(2)}}\right)^2
\end{equation}
(method adapted from Lockman, 1989). In these equations $\Delta\nu$ is the width of the line in frequancy, $\Delta
V_{\rm{FWHM}}$ is the width of the line in velocity, $c$ is the speed
of light, $k$ is the Boltzmann constant and $m$ is the mass of the
atom. For S259 this gives $T_{\rm{e}} \leq 1.3\times10^4$\,K. This
calculation should provide a generous upper limit as the emission line
will also possess a contribution from pressure (collisional)
broadening, the magnitude of which will depend on the density of the
{\sc Hii} region.\\

\begin{figure}
\centerline{\includegraphics[height=9.cm,width=9.cm,angle=0]{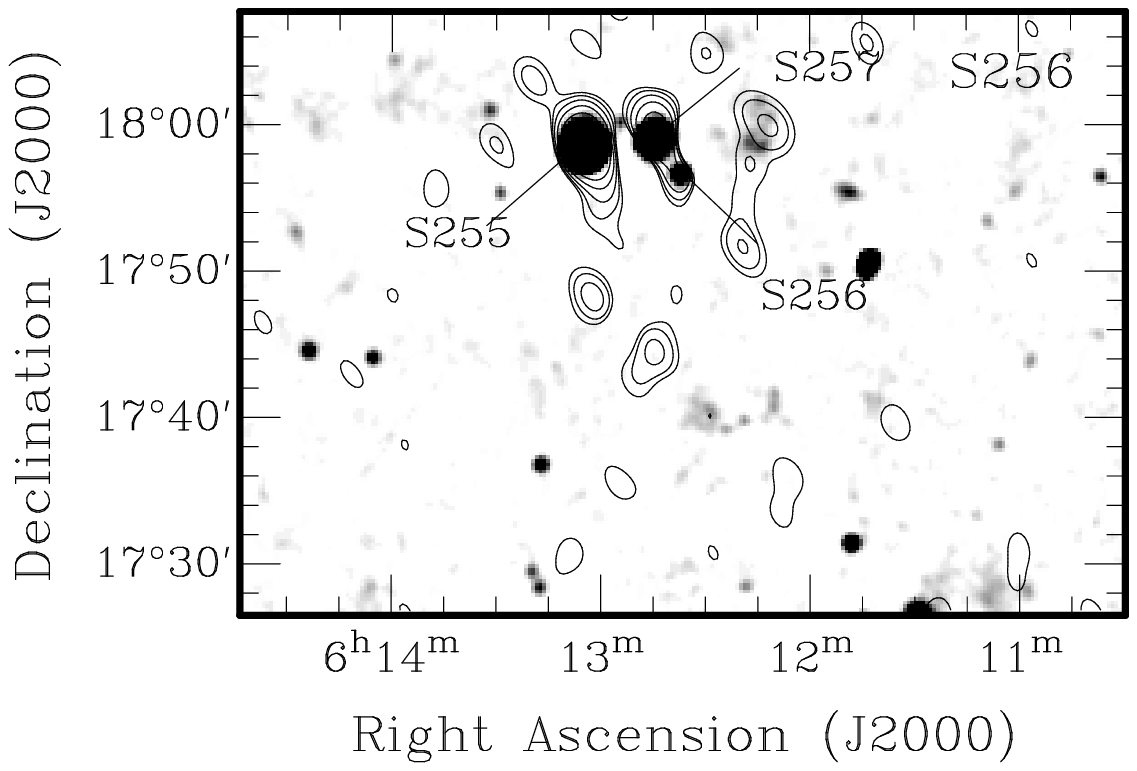}}
\centerline{\includegraphics[height=8.5cm,width=4.cm,angle=-90]{./S256_sp.ps}}
\caption{Above: Map of the S256 region. AMI 15.8\,GHz contours are overlaid
  on an NVSS 1.4\,GHz greyscale image. Contours are as in Figure~\ref{fig:s175spec}. Below: Radio spectrum of S256. Data points are integrated flux
  densities taken from the literature, see
  Table~\ref{tab:radioflux}. The best fitting power law is
  shown as a dashed line. \label{fig:s256spec}}
\end{figure}

\noindent
\textbf{S256} (Figure~\ref{fig:s256spec}) The environment of S256 is complex and the object itself
is dwarfed by its near neighbours S255 and S257. These sources were
excluded from the sample due to their large angular extent and
consequent flux losses for AMI. However, their proximity means that at
the resolution of the AMI our single channel signal-to-noise is not
good enough in this instance to produce separate maps. Therefore we
present only a combined map and flux density for this object.

We estimate an upper limit on the electron gas temperature of S256 using Equation~\ref{equ:te} to find $T_{\rm{e}} \leq$
22\,300\,K. The line width of S256 is broad at 32.3\,km\,s$^{-1}$
(corrected) 
indicating that this is a dense {\sc Hii} region.\\

\begin{figure}
\centerline{\includegraphics[height=9.cm,width=9.5cm,angle=0]{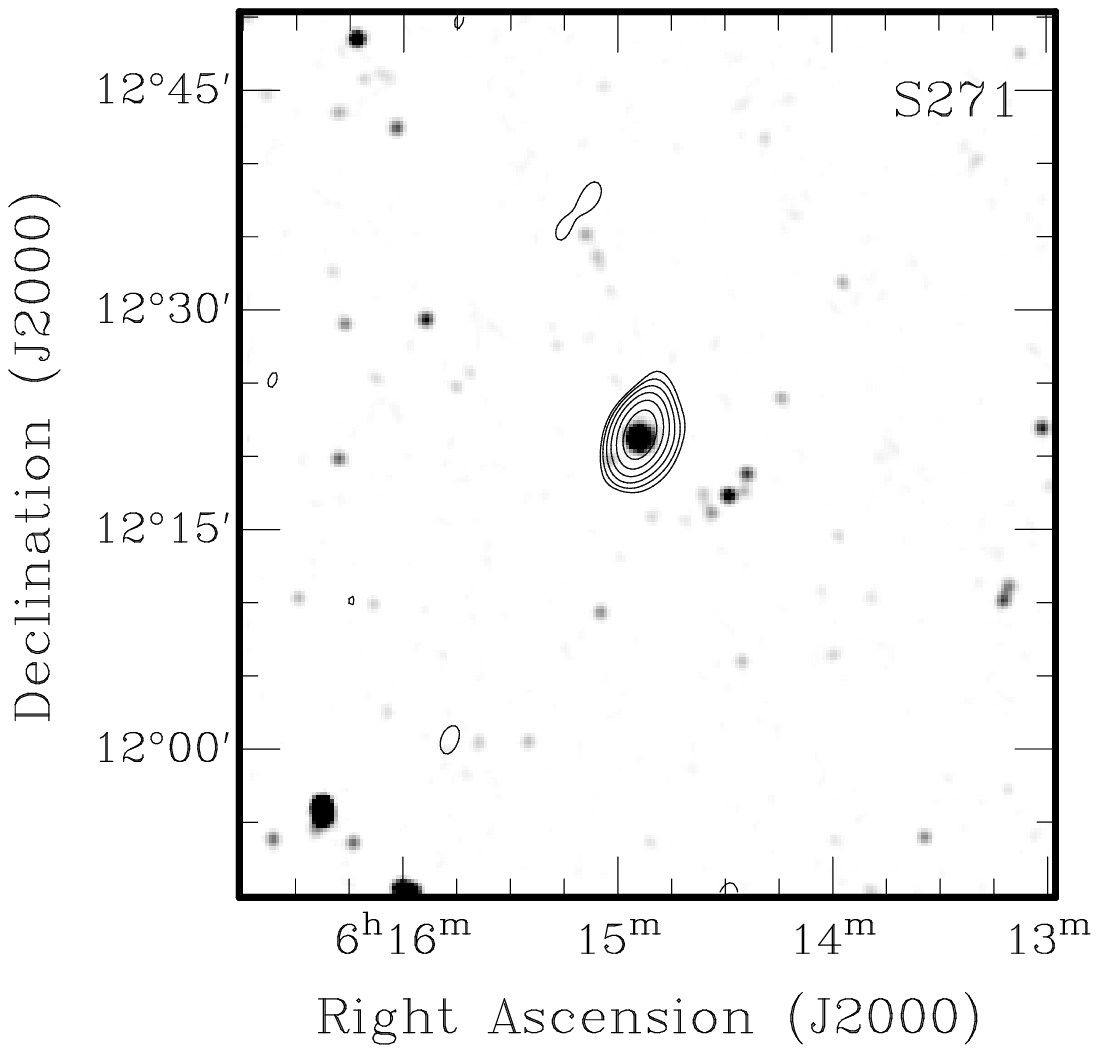}}
\centerline{\includegraphics[height=8.5cm,width=4.cm,angle=-90]{./S271_sp.ps}}
\caption{Above: Map of the S271 region. AMI 15.8\,GHz contours are overlaid
  on an NVSS 1.4\,GHz greyscale image. Contours are as in Figure~\ref{fig:s175spec}. Below: Radio spectrum of S271. Data points are integrated flux
  densities taken from the literature, see
  Table~\ref{tab:radioflux}. The best fitting power law is
  shown as a dashed line. \label{fig:s271spec}}
\end{figure}

\noindent
\textbf{S271} (Figure~\ref{fig:s271spec}) The true nature of S271 is a matter of debate. Although
it was originally thought to be an {\sc Hii} region there is
considerable evidence that it may be in fact a planetary nebula
(PN). This is supported by the IRAS flux densities which show a spike
at 60\,$\mu$m with an excess of $> 100$\,Jy relative to the 12, 25
and 100\,$\mu$m bands. Consequently we cannot fit a dust temperature
for this object but do calculate an electron gas temperature of
$T_{\rm{e}}=7200\pm500$\,K from recombination line data (Hunter 1992). 
In the AMI band this source is heavily contaminated by satellite
interference and although we measure a reasonable spectrum for channels
3 to 5 we then see a turn over in the data. In the absence of any
physical mechanism for this turn over we must conclude that it arises
as a consequence of poor calibration due to satellite
interference. Simulations using the visibility data show that there
should be no significant ($>$\,1\%) flux loss over the AMI band and,
although the phase errors are quite large due to the contaminating
signal, self-calibrating the data has little effect.\\

\begin{figure}
\centerline{\includegraphics[height=9.cm,width=9.cm,angle=0]{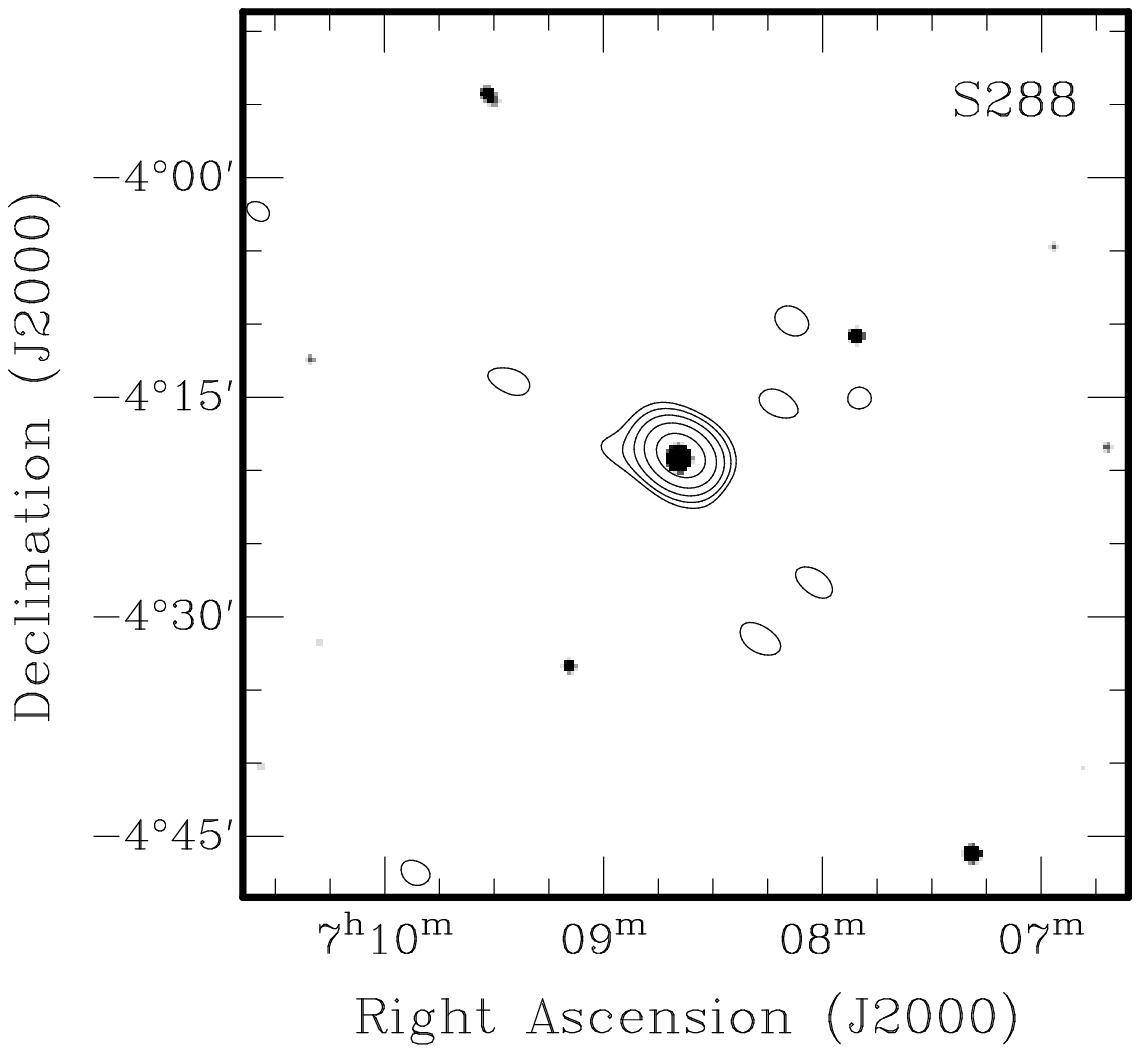}}
\centerline{\includegraphics[height=8.5cm,width=4.cm,angle=-90]{./S288_sp.ps}}
\caption{Above: Map of the S288 region. AMI 15.8\,GHz contours are overlaid
  on an NVSS 1.4\,GHz greyscale image. Contours are as in Figure~\ref{fig:s175spec}. Below: Radio spectrum of S288. Data points are integrated flux
  densities taken from the literature, see
  Table~\ref{tab:radioflux}. The best fitting power law is
  shown as a dashed line. \label{fig:s288spec}}
\end{figure}

\noindent
\textbf{S288} (Figure~\ref{fig:s288spec}) This source has a declination of $-4$\,degrees, right on
the limit of AMI's field of view. At this declination interference
from geostationary satellites contaminates the data severely and
consequently the errors on our data are significantly larger than in
the case of higher declination sources. This is a pity since S288 is a
small bright nebula, comparatively isolated and with a luminous
association seen at 100\,$\mu$m in the IRAS data. In spite of this
contamination we still measure a reasonable microwave spectrum with a
spectral index of $\alpha = 0.18\pm0.06$ across the AMI band. However, we
urge caution due to the poor nature of the data. From the IRAS
100/60\,$\mu$m flux densities we fit a dust temperature of $T_{\rm{d}}
  = 35.4$\,K and from recombination line ratios we calculate a gas
  temperature of $T_{\rm{e}}=6600\pm400$\,K.  \\

\begin{figure}
\centerline{\includegraphics[height=9.cm,width=9.cm,angle=0]{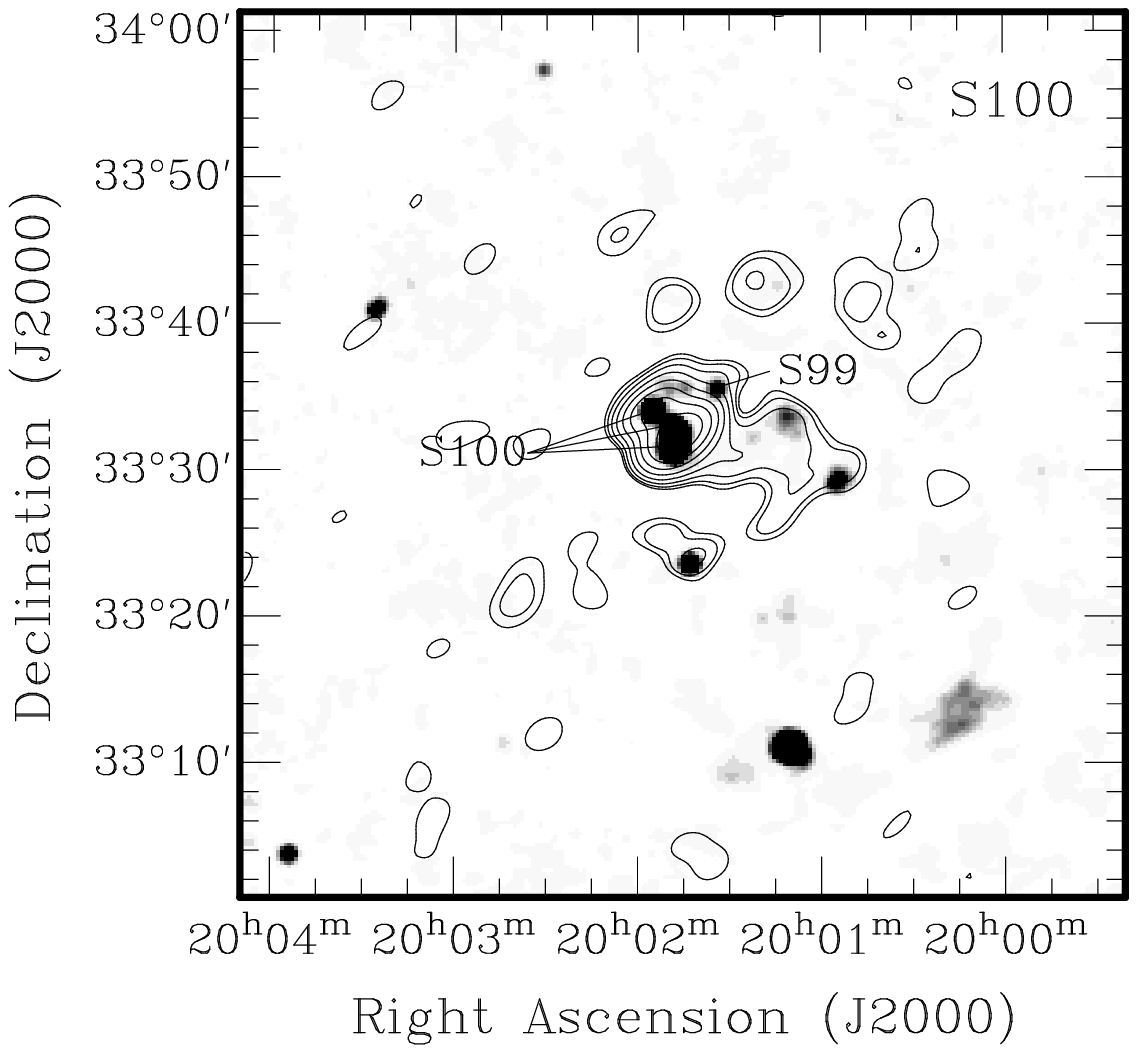}}
\caption{Map of the S\,99/100 region. AMI 15.8\,GHz contours are overlaid
  on an NVSS 1.4\,GHz greyscale image. Contours are as in Figure~\ref{fig:s175spec}. \label{fig:s100map}}
\end{figure}

\noindent
\textbf{S100} (Figure~\ref{fig:s100map}) The S100/99 complex is poorly resolved by the PSF of the
AMI. We measure a peak flux density towards this complex of
$S_{\rm{p}} = 8.28\pm0.41$\,Jy\,bm$^{-1}$. Although we present our combined channel map here we have not
compiled spectra for the different sources since the complicated
nature of the region makes higher resolution necessary for extracting
reliable flux densities.\\

\begin{figure}
\centerline{\includegraphics[height=9.cm,width=9.cm,angle=0]{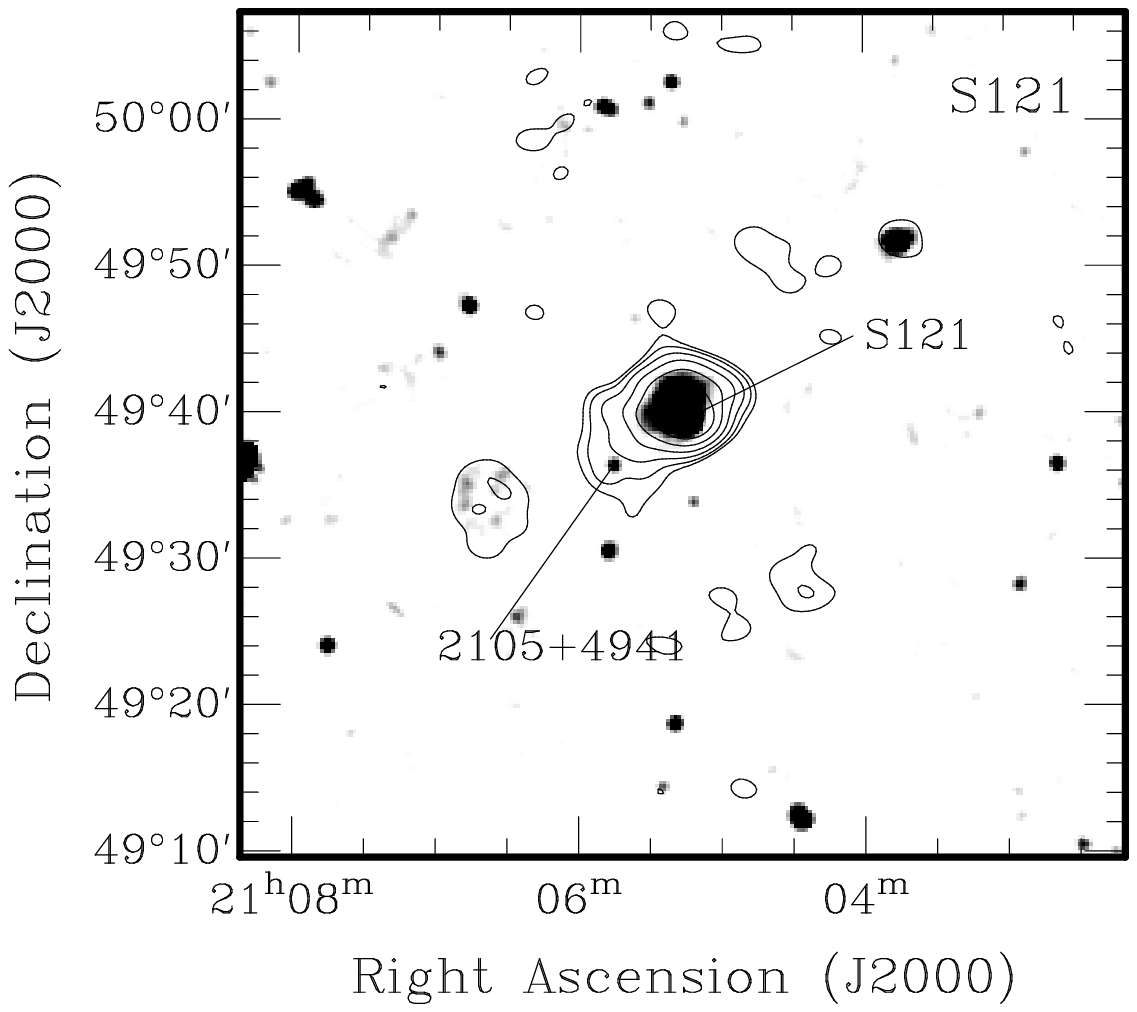}}
\centerline{\includegraphics[height=8.5cm,width=4.cm,angle=-90]{./S121_sp.ps}}
\caption{Above: Map of the S121 region. AMI 15.8\,GHz contours are overlaid
  on an NVSS 1.4\,GHz greyscale image. Contours are as in Figure~\ref{fig:s175spec}. Below: Radio spectrum of S121. Data points are integrated flux
  densities taken from the literature, see
  Table~\ref{tab:radioflux}. Arrows indicate the calculated
  correction for flux losses. The best fitting power law is
  shown as a dashed line. \label{fig:s121spec}}
\end{figure}

\noindent
\textbf{S121} (Figure~\ref{fig:s121spec}) This source is the most extended in our sample
and is poorly fitted by a Gaussian at higher frequencies. Channels 6
and 7 of the AMI data show severe ($>$\,20\%) flux loss towards this object
due to a lack of short spacings caused by necessary flags in the
data. Using the visibility coverage of these channels and the
total power maps of the CGPS we are able to calculate these
losses and their corrected flux densities are indicated in
Figure~\ref{fig:s121spec} by an arrow. The IRAS 100\,$\mu$m emission
closely traces the 15.8\,GHz map with extensions to the north and
south-east.
Fitting a modified Planck spectrum to the IRAS 100/60\,$\mu$m flux
densities gives a dust temperature of $T_{\rm{d}} = 28.8$\,K. Vallee
(1983) derived the electron gas temperature of this region,
$T_{\rm{e}} = 9000$\,K, from observations of the H85$\alpha$ radio
recombination line and measured the emission measure to be
$8\times10^4$\,pc\,cm$^{-6}$. The spectrum of S121 is shown in
Figure~\ref{fig:s121spec}; in addition to VLA data points at 1.4
and 4.89\,GHz (Fich 1993) we include data at 10.5\,GHz taken by Vallee
(1983) with the Alonquin 46\,m dish.\\
\begin{table}
\begin{center}
\caption{Dust and gas temperatures for the AMI {\sc Hii} region sample.\vspace{0.2cm}\label{tab:temp}} 
\begin{tabular}{lcc}
\hline\hline
Name & $T_{\rm{dust}}$ & $T_{\rm{e}}$ \\
     & (K)             & (K)          \\
\hline
\hline 
S175& 27.9 & $7000\pm200$ \\	
S186& 33.2 & 7300 \\	
S211& 19.2 & $7000\pm200$ \\	
BFS46& 38.5 & - \\	
S259& 29.2 & $\leq$13000 \\	 
S256& 18.7 & $\leq$22300 \\	
S271& -    & $7200\pm500$ \\	
S288& 35.4 & $6600\pm400$ \\	
S121& 28.8 & 9000$^{(1)}$ \\	
S127& 30.3 & $10500\pm820$ \\	
BFS10& 29.0 & - \\	 
S138& 35.4 & 6300$^{(2)}$, 11200$^{(3)}$ \\	
S149& 31.4 & $7500\pm400$, $8600\pm495^{(4)}$ \\	
S152& 30.4 & 8400$^{(5)}$, $9100\pm900^{(6)}$ \\	
S167& 33.3 & $\leq$7625 \\	

\hline
\end{tabular}
\begin{minipage}{7cm}
{\small\vspace{0.1cm} Notes:--Temperatures are calculated as described
in the text with the exceptions of (1) Vallee (1983), (2) Deharveng
et~al.(1999), (3),(5) Afflerbach (1997), (4) Matthews (1981), and (6)
Wink (1983). } 
\end{minipage}
\end{center}
\end{table}

\begin{figure}
\centerline{\includegraphics[height=9.cm,width=9.cm,angle=0]{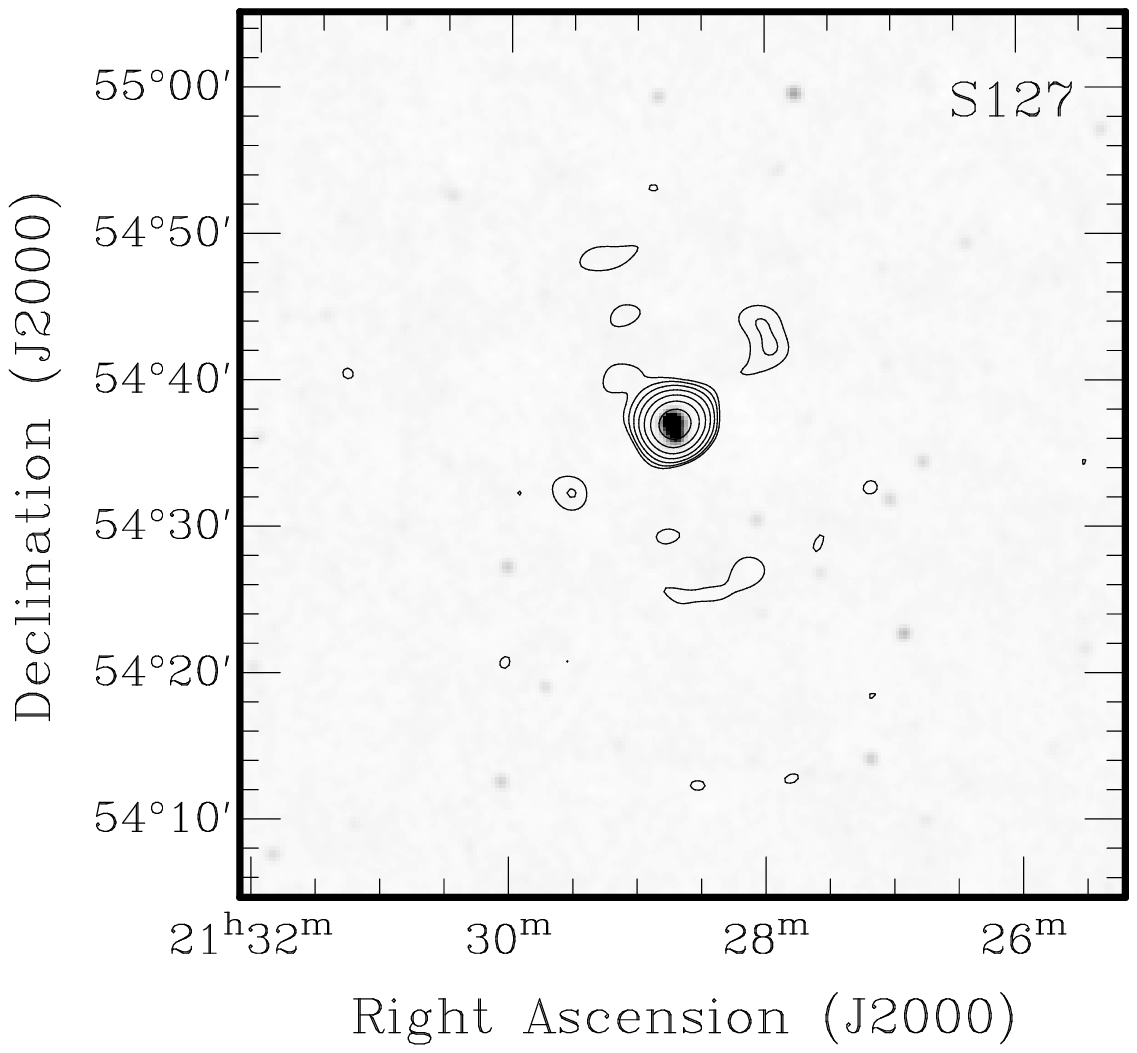}}
\centerline{\includegraphics[height=8.5cm,width=4.cm,angle=-90]{./S127_sp.ps}}
\caption{Above: Map of the S127 region. AMI 15.8\,GHz contours are overlaid
  on an NVSS 1.4\,GHz greyscale image. Contours are as in Figure~\ref{fig:s175spec}. Below: Radio spectrum of S127. Data points are integrated flux
  densities taken from the literature, see
  Table~\ref{tab:radioflux}. In addition data at
  4.86, 8.44
  and 15\,GHz from the VLA are also shown. The best fitting power law is
  shown as a dashed line. \label{fig:s127spec}}
\end{figure}

\noindent
\textbf{S127} (Figure~\ref{fig:s127spec}) Otherwise known as LBN\,436 this region in fact contains two
compact {\sc Hii} regions (WB\,85A and WB\,85B) which are unresolved by the AMI. Indeed at
such small separation the flux density of S127 as found in the
literature frequently comprises both objects. In Figure~\ref{fig:s127spec} we show
data from NVSS at 1.4\,GHz; 2.7\,GHz data from Paladini et al. (2003) re-analysed from the
Effelsberg 100\,m telescope; 4.89\,GHz data (Fich
1993) and 4.86\,GHz (Rudolph et~al. 1996) from the VLA. We also
include data at 8.44\,GHz and 15\,GHz (Rudolph et~al. 1996) from the
VLA, although we do not use these data for fitting purposes as we
believe them to be heavily affected by flux losses. We allow the poorer resolution Effelsberg data in this instance
due to the relatively isolated nature of the complex. 

The IRAS
100\,$\mu$m emission is very similar to that at 15.8\,GHz, with no
secondary sources in the field. Fitting a modified Planck spectrum to the 100 and 60\,$\mu$m data  gives a dust
temperature of 30.3\,K. Significant excess is present at shorter IR
wavelengths indicating the presence of a second hotter dust
component. Although radio recombination line or optical recombination line data are not
available for this object we can estimate the temperature of the
electron gas using its distance from the Galactic centre (Afflerbach,
Churchwell \& Werner 1997).

\begin{equation}
T_{{\rm e}}/{\rm K} = (4560\pm220)+(390\pm40)(D_{{\rm G}}/{\rm kpc}).
\end{equation}

We note that at a distance of $D_{\rm{G}}$=15\,kpc from the Galactic
 center S127 is outside the range of data fitted to derive this
 relation. However, the calculated temperature of 10500$\pm820$\,K is similar
 to that extrapolated from measurements by Fich \& Silkey (1991) which
 suggest a value of 10$^4$\,K in the far outer galaxy.\\

\begin{figure}
\centerline{\includegraphics[height=9.cm,width=9.cm,angle=0]{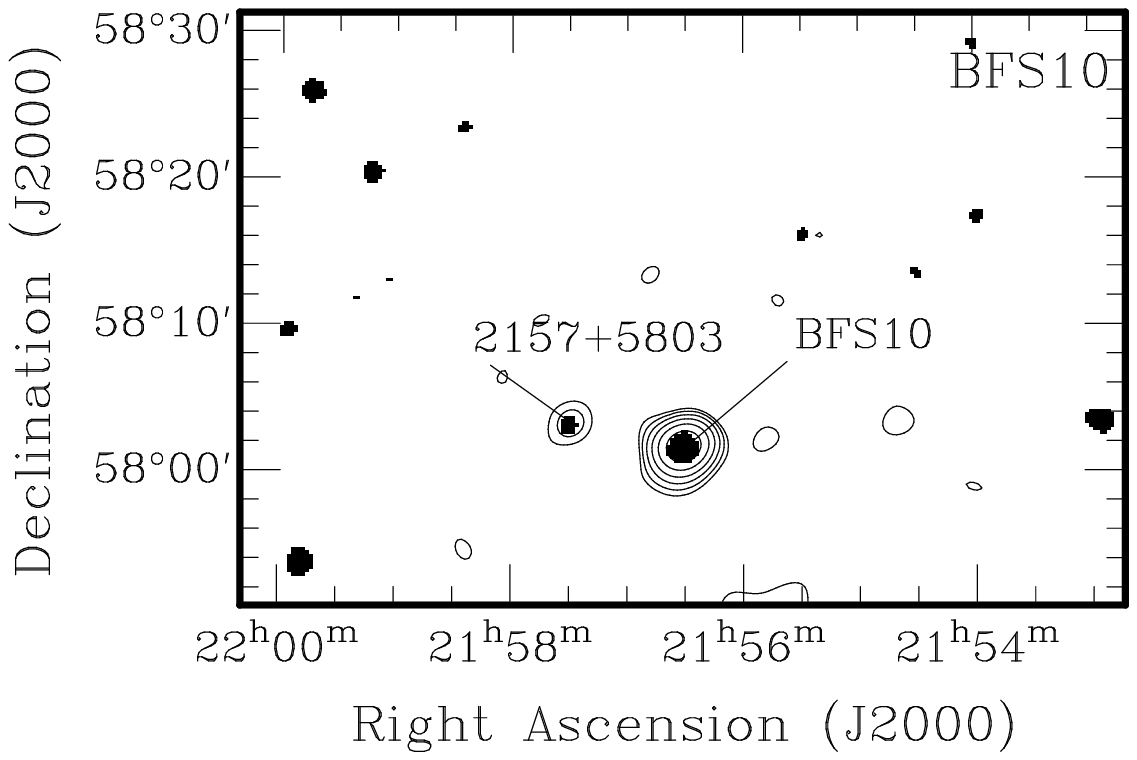}}
\centerline{\includegraphics[height=8.5cm,width=4.cm,angle=-90]{./BFS10_sp.ps}}
\caption{Above: Map of the BFS10 region. AMI 15.8\,GHz contours are overlaid
  on an NVSS 1.4\,GHz greyscale image. Contours are as in Figure~\ref{fig:s175spec}. Below: Radio spectrum of BFS10. Data points are integrated flux
  densities taken from the literature, see
  Table~\ref{tab:radioflux}. The best fitting power law is
  shown as a dashed line. \label{fig:bfs10spec}}
\end{figure}

\noindent
\textbf{BFS10} (Figure~\ref{fig:bfs10spec}) The CO selected {\sc Hii} region BFS10 is a little
studied object. It is dwarfed on larger scales by the neighbouring
{\sc Hii} complexes S131 and DA568. We fit a dust temperature of
$T_{\rm{d}}=29.0$\,K. We see an increased amount of flux in AMI channel
3 towards BFS10. This seems to be caused by an amount of extended
emission associated with this object which is seen on the largest
scales only. This extended emission may also be the cause of the
slight discrepancy between the VLA and GB6 fluxes at 4.8\,GHz. It does
not appear to be immediately obvious in the map plane but is visible
as structure on large angular scales in the visibility data of channel
3; the visibility coverage of which extends to slightly larger angular scales
then channels 4--8 due to its lower frequency.\\

\begin{figure}
\centerline{\includegraphics[height=9.cm,width=9.cm,angle=0]{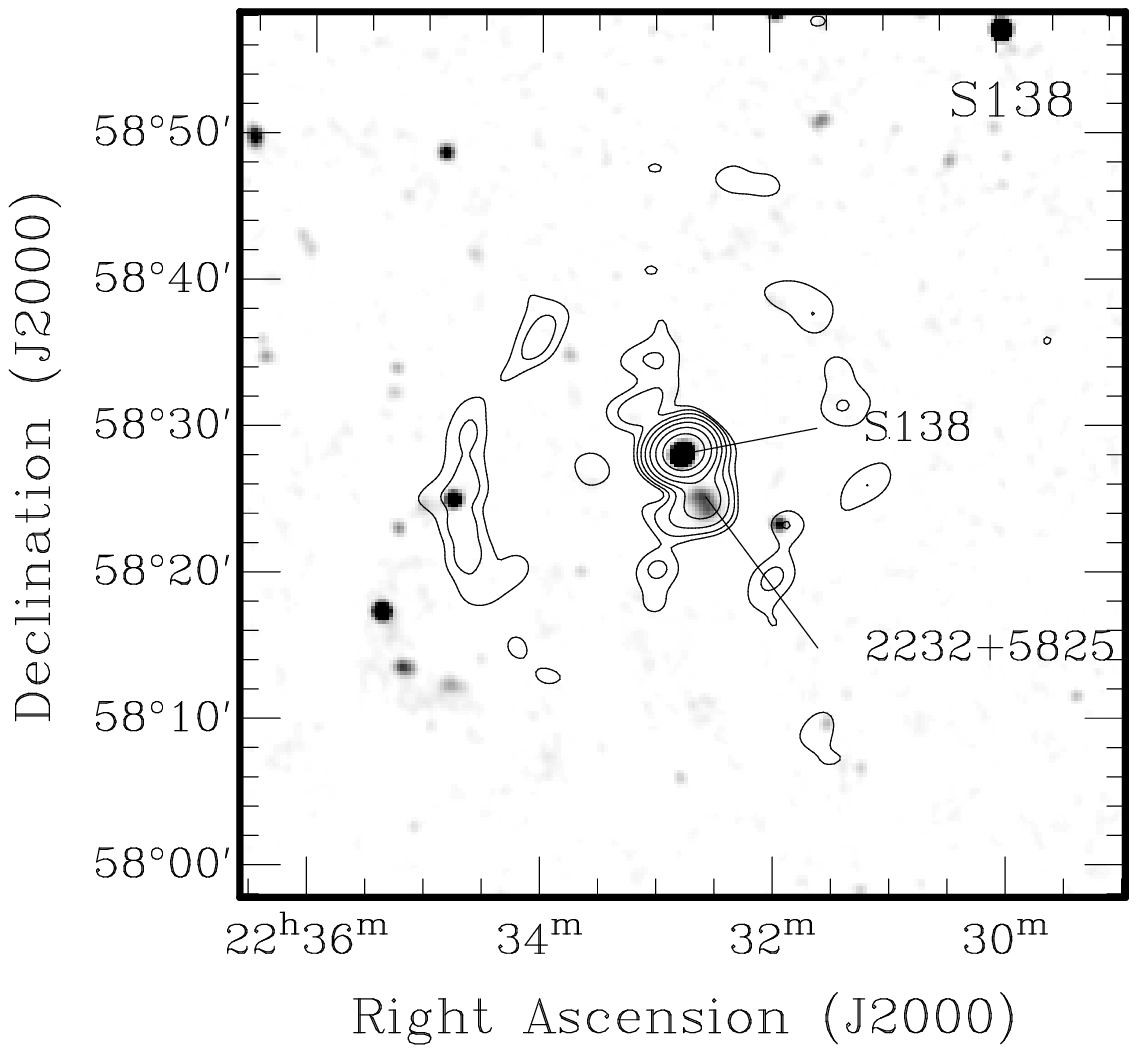}}
\centerline{\includegraphics[height=8.5cm,width=4.cm,angle=-90]{./S138_sp.ps}}
\caption{Above: Map of the S138 region. AMI 15.8\,GHz contours are overlaid
  on an NVSS 1.4\,GHz greyscale image. Contours are as in Figure~\ref{fig:s175spec}. Below: Radio spectrum of S138. Data points are integrated flux
  densities taken from the literature, see
  Table~\ref{tab:radioflux}. The best fitting power law is
  shown as a dashed line. \label{fig:s138spec}}
\end{figure}

\noindent
\textbf{S138} (Figure~\ref{fig:s138spec}) This {\sc Hii} region, associated with IRAS\,22308+5812,
is in a complex region at
both radio and infra-red wavelengths. In the radio the flux density
measurement of the GB6 survey at 4.85\,GHz is confused by the nearby
source NVSS 2232+5832. The 100$\mu$m IRAS data shows a broad band of
diffuse emission around S138 and a neighbouring source to the
south-west. The radio continuum data at 15.8\,GHz also shows an
extension to the south-west consistent with the position of both the
radio and IR source (IRAS\,22306+5809). From the {\sc Nii} and
H$\alpha$ spectral measurements of Deharveng et~al. (1999) we can
derive an electron gas temperature for the region, using
Equation~\ref{equ:nh}. The value of $T_{\rm{e}} = 6300$\,K that we
  calculate is somewhat lower than that of Afflerbach et~al. (1997)
  who derive a temperature of 11200\,K using the ratio of infra-red
  fine structure lines to the radio continuum at lower frequencies. We
  suggest that this may be a consequence of the optical recombination
  line measurements being taken in the direction of the main exciting
  star where the electron density is highest (N$_{\rm{e}} \simeq
  1000$\,cm$^{-3}$), compared to the outer regions of the nebula where
  it falls to $\sim$200\,cm$^{-3}$ (Deharveng 1999), which is more
  in line with the value calculated by Afflerbach of
  175\,cm$^{-3}$. From the IRAS 100/60\,$\mu$m flux densities we
  calculate a dust temperature, $T_{\rm{d}} = 35.4$\,K.\\

\begin{figure}
\centerline{\includegraphics[height=9.cm,width=9.cm,angle=0]{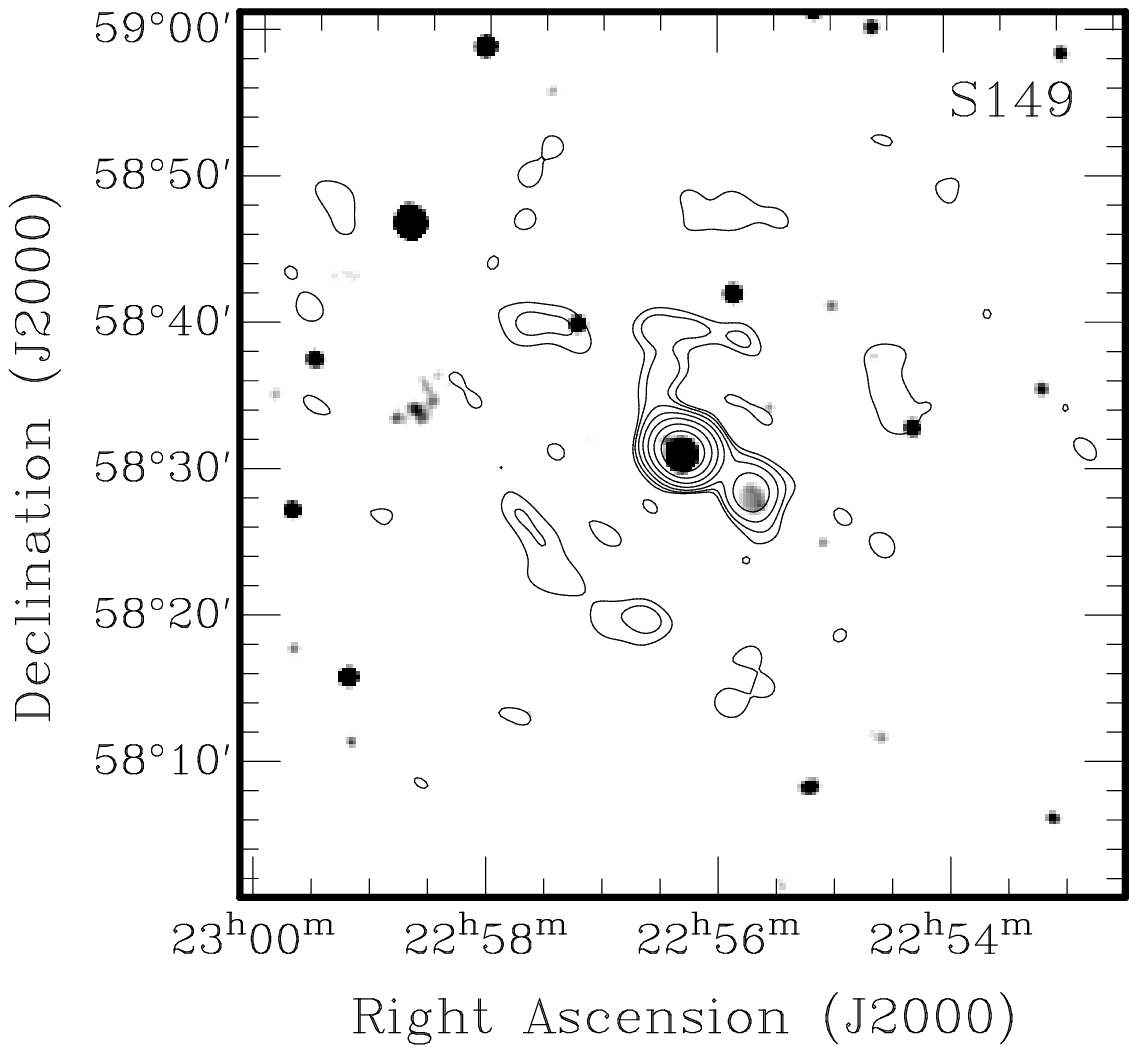}}
\centerline{\includegraphics[height=8.5cm,width=4.cm,angle=-90]{./S149_sp.ps}}
\caption{Above: Map of the S149 region. AMI 15.8\,GHz contours are overlaid
  on an NVSS 1.4\,GHz greyscale image. Contours are as in Figure~\ref{fig:s175spec}. Below: Radio spectrum of S149. Data points are integrated flux
  densities taken from the literature, see
  Table~\ref{tab:radioflux}. The best fitting power law is
  shown as a dashed line. \label{fig:s149spec}}
\end{figure}

\noindent
\textbf{S149} (Figure~\ref{fig:s149spec}) The source we identify with S149 is in reality both S149
and S148, whose small separation makes them indistinguishable at the
resolution of AMI. The S147 {\sc Hii} region is also visible as an
extension to the south-west of the main source. All three sources are
visible in the IRAS 100\,$\mu$m data as is a diffuse extension to the
north of the main source, which possesses no radio
counterpart. Fitting to the IRAS 100/60\,$\mu$m flux densities we
find a dust temperature of $T_{\rm{d}} = 31.4$\,K for S149; and using
Equation~\ref{equ:nh} and the optical recombination line data of
Hunter et~al. (1992) we find an electron gas temperature of
$T_{\rm{e}} = 7500\pm400$\,K, slightly lower than that of Matthews (1981)
who found $T_{\rm{e}} = 8600\pm495$\,K.\\

\begin{figure}
\centerline{\includegraphics[height=9.cm,width=9.cm,angle=0]{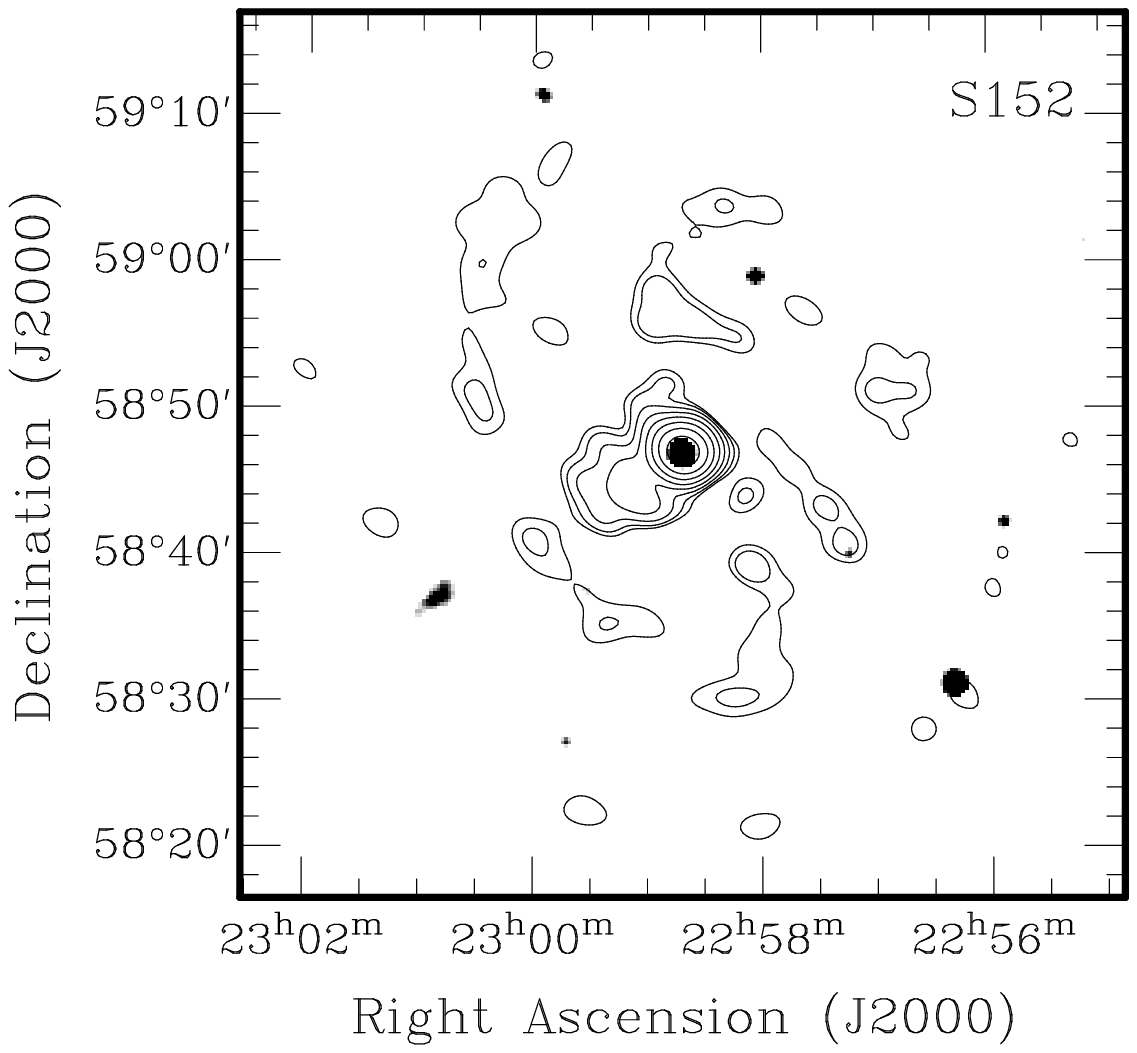}}
\centerline{\includegraphics[height=8.5cm,width=4.cm,angle=-90]{./S152_sp.ps}}
\caption{Above: Map of the S152 region. AMI 15.8\,GHz contours are overlaid
  on an NVSS 1.4\,GHz greyscale image. Contours are as in Figure~\ref{fig:s175spec}. Below: Radio spectrum of S152. Data points are integrated flux
  densities taken from the literature, see
  Table~\ref{tab:radioflux}. In addition data at 8.35\,GHz
  from the Galactic plane survey of Langston (2000) are also shown. The best fitting power law is
  shown as a dashed line. \label{fig:s152spec}}
\end{figure}

\noindent
\textbf{S152} (Figure~\ref{fig:s152spec}) The radio emission from S152 appears slightly offset from
the IRAS 100\,$\mu$m peak. There is a large diffuse patch of radio
emission to the south-east of the object, see Figure~\ref{fig:s152spec}, which has no corresponding IR
counterpart although there are several IR sources nearby. The electron
gas temperature calculated by Afflerbach et~al. (1997) of $T_{\rm{e}} =
  8400$\,K agrees well with that of Wink et~al. (1983) who find
  $T_{\rm{e}} = 9100\pm900$\,K using the RRL H76$\alpha$. \\

\begin{figure}
\centerline{\includegraphics[height=9.cm,width=9.cm,angle=0]{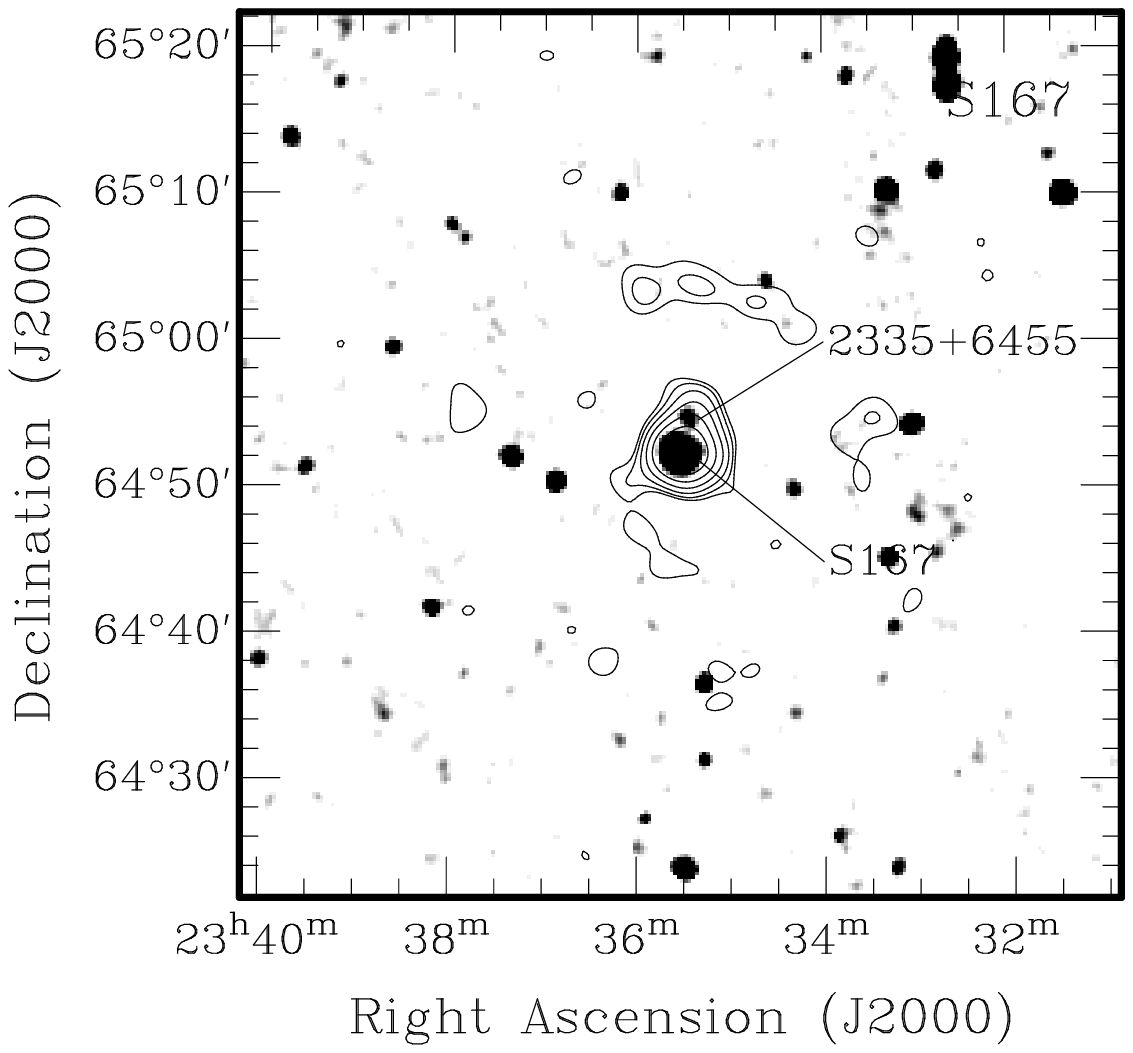}}
\centerline{\includegraphics[height=8.5cm,width=4.cm,angle=-90]{./S167_sp.ps}}
\caption{Above: Map of the S167 region. AMI 15.8\,GHz contours are overlaid
  on an NVSS 1.4\,GHz greyscale image. Contours are as in Figure~\ref{fig:s175spec}. Below: Radio spectrum of S167. Data points are integrated flux
  densities taken from the literature, see
  Table~\ref{tab:radioflux}. The best fitting power law is
  shown as a dashed line. \label{fig:s167spec}}
\end{figure}

\noindent
\textbf{S167} (Figure~\ref{fig:s167spec}) Although originally classified as a planetary nebula
further investigation confirmed the status of S167 as an {\sc Hii}
region (Acker 1990). At 15.8\,GHz we see a relatively compact source
with slight extensions to the north and south-east, the first of these
coinciding with the radio source NVSS 2335+6455. We observe a
relatively constant spectral index from 1.4 to 18\,GHz with
$\alpha_{\rm{AMI}} = 0.16\pm0.54$ and an overall index of
$\alpha_{18}^{1.4} = 0.13\pm0.06$.

As with S256 and S259 we can only place an upper limit on the
electron gas temperature of S167 and find, using
Equation~\ref{equ:te}, $T_{\rm{e}} \leq$ 7625\,K.\\

\begin{table*}
\centering
\caption{AMI {\sc Hii} sample \vspace{0.2cm}\label{tab:spectra}} 
\begin{tabular}{lccccccc}
\hline\hline
Name & S$^{15}_{\rm{i}}$ & $\alpha_{1.4}^{4.9}$&
$\alpha_{\rm{AMI}}$&$\alpha_{\rm{tot}}$&S$_{\rm{excess}}$&Excess
100$\mu$m emissivity\\
&(Jy)&&&&(mJy)&$\mu$K(MJy/sr)$^{-1}$\\
&[2]&[3]&[4]&[5]&[6]&[7]\\
\hline
\hline 
S175& $0.095\pm0.005$ & -0.15$\pm$0.36 & 0.51$\pm$2.82 &
0.04$\pm$0.07 & $<83$ & $<60$\\	
       &                 & 0.10           &               &
&$<11$& $<6$\\ 
S186& $0.144\pm0.007$ & 0.02$\pm$0.09 & 0.27$\pm$1.34 &
0.07$\pm$0.06 & $<21$ & $<11$\\	
       &                 & 0.10           &               &
&$<29$& $<15$\\ 
S211& $0.568\pm0.028$ & 0.20$\pm$0.07 & 0.06$\pm$0.05 &
0.14$\pm$0.02& 25$^{+47}_{-45}$& $13$\\	
       &                 & 0.10           &               &
& $<114$ & $<60$ \\ 
BFS46& $0.201\pm0.010$ & 0.02$\pm$0.06 & 0.12$\pm$0.81 & 0.09$\pm$0.04 & $<1$ &$<1$\\	
       &                 & 0.10           &               &
&$<28$& $<6$\\ 
S259$^{a}$...& $0.071\pm0.004$ & -0.03$\pm$0.13 & 0.48$\pm$5.66 &
0.29$\pm$0.29 & - & -\\	 
       &                 & 0.10           &               &               &$<1$&$<1$\\ 
S256$^{a}$...& $0.094\pm0.010$ & 0.26$\pm$0.55 & - & 0.21$\pm$0.31 &
9$^{+75}_{-41}$ &3 \\	
       &                 & 0.10           &               &
&$<21$& $<6$\\ 
S271$^{a,b}$...& $0.185\pm0.020$ & 0.08$\pm$0.11 & 1.01$\pm$5.67 & 0.27$\pm$0.23 & $<1$ &$<1$\\	
       &                 & 0.10           &               &               &$<1$&$<1$\\ 
S288$^{a}$...& $0.440\pm0.044$& 0.06$\pm$0.04 & 0.25$\pm$1.41 &
0.18$\pm$0.06 & $<1$ & $<1$ \\	
       &                 & 0.10           &               &
& $<46$ & $<23$ \\ 
S121& $0.440\pm0.022$ & 0.02$\pm$0.05 & -0.08$\pm$0.26 &
0.07$\pm$0.01 & $1^{+37}_{-35}$ & $<1$ \\	
       &                 & 0.10           &               &
& $7^{+44}_{-44}$ &8 \\ 
S127& $0.415\pm0.021$ & 0.07$\pm$0.05 & 0.19$\pm$0.41 &
0.16$\pm$0.02 &$<1$ & $<1$ \\	
       &                 & 0.10           &               & & $<20$ &
$<19$ \\ 
BFS10& $0.135\pm0.007$ & 0.08$\pm$0.16 & 0.38$\pm$2.05 & 0.19$\pm$0.09 & $<32$ &$<39$\\	 
       &                 & 0.10           &               &               &$<2$&$<2$\\ 
S138& $0.394\pm0.020$ & -0.03$\pm$0.03 & 0.25$\pm$0.51 &
0.13$\pm$0.03 & - & - \\	
       &                 & 0.10           &               &
& $<6$ & $<2$ \\ 
S149& $0.410\pm0.021$ & 0.09$\pm$0.07 & 0.25$\pm$0.39 & 0.14$\pm$0.01
& $<29$ & $<20$ \\	
       &                 & 0.10           &               &
& $<53$ & $<34$\\ 
S152& $1.048\pm0.052$ & -0.04$\pm$0.06 & 0.17$\pm$0.10 &
0.11$\pm$0.01 & - & - \\
       &                 & 0.10           &               &
& $<81$ & $<16$ \\ 
S167& $0.175\pm0.009$ & 0.08$\pm$0.14 & 0.16$\pm$0.54 & 0.13$\pm$0.06
&$<48$ & $<42$\\	
       &                 & 0.10           &               &
&$<23$& $<23$\\ 

\hline
\end{tabular}
\begin{minipage}{16cm}
{\small\vspace{0.1cm} Notes:-- [2] 15.8\,GHz integrated flux densities, [3]
  spectral index calculated from 1.4 and 4.89\,GHz VLA data; [4]
  spectral index fitted to the 6 spectral channels of the AMI, [5]
  overall spectral index including both VLA and AMI data; [6] derived
  excess emission at 15.8\,GHz, upper limits are at $2\sigma$, $\sim$95\%; [7]
  excess 100\,$\mu$m emissivity following the method of Dickinson
  et~al. (2007). (a) Significant satellite interference;
(b) Planetary Nebula\label{tab:results}\\} 
\end{minipage}
\end{table*}

\section{Discussion}

{\sc Hii} regions are a reasonable place to look for anomalous emission
since their general radio behaviour is well understood. In the
region of the spectrum above approximately 1\,GHz they are
dominated by thermal free-free emission with a canonical spectral
index of $\alpha = 0.1$. In an idealized sense this emission arises
from a sphere of ionized gas surrounding a hot star, or cluster of
stars; although the {\sc Hii} region itself may consist of several compact
objects which are unresolved by the synthesized beam of the AMI. In
addition to this, {\sc Hii} regions are strong emitters in the IR
making them suitable candidates for spinning dust emission, and have 
 dust temperatures typically in the range 30--50\,K. All the objects
 presented here have temperatures consistent with this range, see Table~\ref{tab:temp}, with the
 exception of S211 which has a slightly lower dust temperature. These objects
 are not necessarily perfect blackbodies and detailed modelling of the
 dust properties, the geometry and the possible spectral features in
 the IRAS bands is beyond the scope of this paper. For more detailed analysis of the
 infrared content of {\sc Hii} regions see, for example, Akabane
 \& Kuno (2005).

The spectrum of optically thin thermal emission varies slowly
with frequency and electron gas temperature, but in the frequency
range used here it can be described well by
a single index of $\alpha = 0.1$. Indeed from our
sample of {\sc Hii} regions we find an average spectral
index between 1.4 and 5\,GHz of $\alpha =0.05\pm0.09$, which is
consistent with this value. Where possible we have calculated the electron gas temperature
of each {\sc Hii} region from data available in the literature. We
find that for all those sources where data are 
available the electron gas temperature of falls within
the expected range (see Table.~\ref{tab:temp}). Furthermore we also find that the flux densities measured at
15.8\,GHz by the AMI are also consistent with this
index. 

Within the AMI band the average spectral index tends to be steeper,
$\alpha = 0.29\pm0.25$, but is not significantly different from the
canonical index of $\alpha = 0.1$. Overall we find an average spectral index of $\alpha^{17.6}_{1.4} =
0.15\pm0.07$.  This confirms the dominance of free--free emission in
these bright 
{\sc Hii} regions.

The aim of this study was to investigate a possible excess of emission
within the AMI band which might be attributed to a spinning dust
component. The AMI is particularly suitable for this type of
measurement since the spinning dust predictions of Drain \& Lazarian
(1998a,b) imply that the peak of the resulting spectrum will lie close to
15\,GHz. In spite of this we see no evidence for an excess in the
sources we observe here. In Table~\ref{tab:results} we present the
combined channel flux density of each source at 15.8\,GHz, the
spectral index from the VLA radio measurements at 1.4 and 4.89\,GHz,
the spectral index calculated using only the data from the six AMI
channels, and the spectral index found using both VLA and AMI data. We
note that the uncertainties on the spectral indices calculated using AMI data
alone include errors which are correlated between the channels, and
that this leads to an overestimation of the uncertainty in each 
spectral index. The average difference between the spectral index
calculated within the AMI band and the overall index from 1.4 to
17.9\,GHz is 0.19 and this is perhaps a better representation of the
non-systematic error in this quantity. In
addition to these derived quantities we also calculate the
{\it excess} towards each {\sc Hii} region at 15.8\,GHz. We do this in
two ways: firstly by extrapolating the spectral index calculated from
the VLA data, $\alpha_{1.4}^{4.9}$; secondly using the canonical
spectral index of $\alpha = 0.1$. Column 6 of Table~\ref{tab:results}
shows the calculated excess at 15.8\,GHz relative to these two
indices. In the case that a positive excess is not found and instead
there is a decrement at 15.8\,GHz towards an object then the 2\,$\sigma$
upper limit is shown. In the case that the 2\,$\sigma$ upper limit is
still a decrement with respect to the flux density predicted from the
spectral index then the excess is marked $<1$. From
Table~\ref{tab:results} it can be seen that a positive difference in
flux density with respect to that predicted from the extrapolated
spectral index is seen only in four of the fifteen {\sc Hii} regions
tabulated here; but that in each of these instances the {\it excess} is not
significant at even the 1\,$\sigma$ level.

Combining the predictions we see that the average excess towards this
sample of fifteen {\sc Hii} regions is $-75$\,mJy extrapolating the
derived spectral index. Using a canonical index of $\alpha= 0.1$ we
see an average excess of $-49$\,mJy. These results would suggest that,
not only is there no evidence for anomalous emission in the spectra of
these objects, but also that there is a slight steepening of the spectral
index as we move to higher frequencies (Dickinson
et~al. 2003). 

This result differs to that of Dickinson et~al. (2007) who found
a slight excess of emission at 31\,GHz for a sample of southern {\sc
  Hii} regions. In terms of the physical characteristics
(dust/electron temperature) the two samples are similar. Observationally the measurements of Dickinson et~al. were made for a
range of slightly larger angular scales, and although we have
shown flux losses not to be significant, 
this only relates to the free--free emission and not to any possible
anomalous component. It has been suggested that the anomalous emission 
is distributed more diffusely (de Olivera--Costa et~al. 2002), and is
consequently affected to a larger degree. However, given the compact
nature of these objects this seems unlikely.

All the objects observed here have bright dust associations and we
investigated the correlation of the 15.8\,GHz flux with that found in
each of the 12, 25, 60 and 100\,$\mu$m bands of IRAS. After
omitting two outlying objects, namely BFS46
and S256, the fluxes of which are artificially high in the IRAS data
due to resolution effects, we performed a Pearson correlation
analysis. We find a positive correlation for all four IRAS bands with
the strongest correlation occuring in the 100\,$\mu$m band ($r$ =
0.88). The correlation between the AMI 15.8\,GHz flux
densities and those of the VLA at 1.4\,GHz is much stronger ($r$ = 0.99),
suggesting that the emission we see at 15.8\,GHz is indeed simply 
free--free rather than dust emission. We note, however, that 
dust emission will depend heavily on the dust conditions
(i.e. temperature, density) within the cloud, and any variance in these
conditions would reduce the degree of correlation.

The limits on any excess have been converted to a dust
emissivity relative to the IRAS 100\,$\mu$m map, see Column 7 of Table~\ref{tab:results}. This conversion is
model independent (Dickinson et~al. 2003) and although we have used
IRAS here other authors have also calculated dust emissivites relative
to different standards such as the DIRBE 140\,$\mu$m map, the Schlegel,
Finkbeiner \& Davis (1998) 100$\mu$m map, or the Finkbeiner, Davis \&
Schlegel (1999) model 8 map normalized at 94\,GHz. Our three positive
differences in flux density: S211, S121 and S256, are all consistent with the expected
emissivity value of $\sim 10\mu \rm{K} (\rm{MJy/sr})^{-1}$ at high
latitudes (Davies et~al. 2006).

\section{Conclusions}

The observations of fifteen bright {\sc Hii} regions and one planetary nebula reported here show no
evidence for anomalous emission due to spinning dust at six
frequencies between 14 and 18\,GHz. This result confirms the dominance
of free-free emission in 
these objects with a spectral index consistent with the canonical
value of 0.1. No significant evidence for spinning dust emission has
been found.

\section{ACKNOWLEDGEMENTS} 

We thank the staff of the Lord's Bridge observatory for their
invaluable assistance in the commissioning and operation of the
Arcminute Microkelvin Imager. The AMI is supported by the STFC. NHW and MLD 
acknowledge the support of a PPARC studentship. We also thank the
anonymous referee for their useful comments.

\begin{table*}
\caption{Radio flux densities. \vspace{0.2cm}\label{tab:radioflux}} 
\begin{tabular}{lcccccccc}
\hline\hline
     &\multicolumn{8}{c}{Freq. (GHz)}\\
     \cline{2-9}
     & 1.4$^{(1)}$& 2.7$^{(2)}$&3.2$^{(3)}$&4.85$^{(4)}$&4.89$^{(5)}$&6.6$^{(3)}$&8.45&10.7$^{(3)}$\\
Name &(Jy)&(Jy)&(Jy)&(Jy)&(Jy)&(Jy)&(Jy)&(Jy)\\
\hline
\hline 
S175...&0.102$\pm$0.004 &0.11&0.120$\pm$0.011&-&0.120&0.123$\pm$0.012&-&-\\	
S186...&0.169$\pm$0.007 &0.18&-&0.169$\pm$0.015&0.165&-&-&-\\	
S211...&0.800$\pm$0.025 &0.74        &0.75$\pm0.10$&0.627$\pm0.056$&0.652&0.6$\pm$0.01&-&0.6$\pm$0.01\\		
BFS46..&0.253$\pm$0.008 &-&-&0.213$\pm$0.021&0.248&-&0.248$\pm$0.012$^{(7)}$&-\\	S259...&0.132$\pm$0.005 &-&-&0.146$\pm$0.015&0.136&-&-&-\\	 
S256...&0.158$\pm$0.016 &-&-&-&0.114&-&-&-\\	
S271...&0.330$\pm$0.011 &0.32&-&0.290$\pm$0.026&0.297&-&-&-\\	
S288...&0.638$\pm$0.024 &0.68&0.90$\pm$0.15&-&0.595&0.6$\pm$0.1&-&0.5$\pm$0.1\\	S121...&0.539$\pm$0.011 &-&-&0.484$\pm$0.043&-&-&-&-\\	
S127...&0.602$\pm$0.010 &0.6$\pm$0.2$^{(8)}$ &-&0.577$\pm0.058$&0.563&-&0.267$\pm$0.026$^{(6)}$&-\\
BFS10..&0.210$\pm$0.007 &-&-&0.176$\pm$0.016&0.190&-&-&-\\	 
S138...&0.537$\pm$0.019 &-&-&0.470$\pm$0.042&0.554&-&-&-\\	
S149...&0.575$\pm$0.019 &-&-&0.545$\pm$0.048&0.512&-&-&-\\	

S152...&1.320$\pm$0.050 &1.56        &-&1.360$\pm$0.140&1.380&-&-&-\\
S167...&0.233$\pm$0.008 &-&-&0.196$\pm$0.017&0.212&-&-&-\\

\hline
\end{tabular}
\normalsize
\begin{minipage}{16cm}
{\small\vspace{0.1cm} Notes:-- Radio flux densities from the literature. Where no error is quoted an
  uncertainty of 10 per cent has been assumed.\\
References:-- (1) Condon et~al. (1998), (2) F{\"u}rst et~al. (1990), (3)
  Andrew et~al. (1973), (4) Gregory (1996), (5) Fich
  (1993), (6) 8.44\,GHz Rudolph et~al. (1996), (7) 8.45\,GHz Felli
  et~al (2006), (8) Paladini et~al. 2003)} 
\end{minipage}
\end{table*}
\begin{table*}
\caption{AMI flux densities. \vspace{0.2cm}\label{tab:amiflux}} 
\begin{tabular}{lcccccc}
\hline\hline
     &\multicolumn{6}{c}{Freq. (GHz)}\\
     \cline{2-7}
     & 14.2& 15.0&15.7&16.4&17.1&17.9\\
Name &(Jy)&(Jy)&(Jy)&(Jy)&(Jy)&(Jy)\\
\hline
\hline 
S175...& 0.103$\pm$0.005&0.098$\pm$0.005&0.095$\pm$0.005&0.092$\pm$0.005&0.092$\pm$0.005&0.091$\pm$0.005\\	
S186...& 0.147$\pm$0.008&0.146$\pm$0.007&0.144$\pm$0.007&0.140$\pm$0.007&0.139$\pm$0.007&0.139$\pm$0.007\\	
S211...& 0.567$\pm$0.025&0.559$\pm$0.025&0.568$\pm$0.025&0.571$\pm$0.025&0.559$\pm$0.025&0.566$\pm$0.025\\	
BFS46..& 0.209$\pm$0.010&0.206$\pm$0.010&0.201$\pm$0.010&0.209$\pm$0.010&0.205$\pm$0.010&0.200$\pm$0.010\\
S259...& 0.075$\pm$0.005&0.068$\pm$0.005&0.071$\pm$0.005&0.065$\pm$0.005&0.068$\pm$0.005&0.065$\pm$0.005\\	 
S256...& -&-&0.094$\pm$0.010&-&-&-\\	
S271...& 0.195$\pm$0.020&0.185$\pm$0.020&0.185$\pm$0.020&0.176$\pm$0.020&0.163$\pm$0.020&0.151$\pm$0.020\\	
S288...& 0.434$\pm$0.043&0.412$\pm$0.041&0.440$\pm$0.044&0.420$\pm$0.042&0.456$\pm$0.046&0.385$\pm$0.039\\	
S121...& 0.453$\pm$0.026&0.435$\pm$0.026&0.440$\pm$0.025&0.461$\pm$0.023&0.380$\pm$0.022&0.332$\pm$0.020\\	
S127...& 0.427$\pm$0.021&0.422$\pm$0.021&0.415$\pm$0.021&0.411$\pm$0.021&0.405$\pm$0.020&0.408$\pm$0.020\\	
BFS10..& 0.149$\pm$0.007&0.138$\pm$0.007&0.135$\pm$0.007&0.133$\pm$0.006&0.132$\pm$0.006&0.128$\pm$0.006\\
S138...& 0.411$\pm$0.020&0.406$\pm$0.020&0.394$\pm$0.020&0.402$\pm$0.020&0.387$\pm$0.020&0.387$\pm$0.020\\	
S149...& 0.424$\pm$0.020&0.418$\pm$0.020&0.410$\pm$0.020&0.408$\pm$0.020&0.392$\pm$0.020&0.407$\pm$0.020\\	
S152...& 1.065$\pm$0.050&1.057$\pm$0.050&1.048$\pm$0.050&1.041$\pm$0.050&1.033$\pm$0.050&1.026$\pm$0.050\\	
S167...& 0.164$\pm$0.008&0.165$\pm$0.008&0.164$\pm$0.008&0.165$\pm$0.008&0.149$\pm$0.007&0.154$\pm$0.008\\	
\hline
\end{tabular}
\normalsize
\begin{minipage}{14cm}
{\small\vspace{0.1cm} Notes:-- Flux densities measured by the AMI
  telescope. No correction has been made for flux losses. See text for
  details.} 
\end{minipage}
\end{table*}

\bsp
\label{lastpage}

\end{document}